\documentclass[aps,prb,twocolumn,floatfix,showpacs,showkeys]{revtex4}

\usepackage{graphicx}
\usepackage{dcolumn} 
\usepackage{amsmath}
\usepackage{bm}      
\usepackage{color}
\usepackage{euscript}
\usepackage{epsfig,psfrag,subfigure,subeqnarray}
\usepackage{bbm}
\usepackage{enumitem}
\def\lan{\langle}
\def\ran{\rangle}
\def\va{\varepsilon}

\def\vk{{\bf k}}

\def\vr{{\bf r}}
\def\vu{{\bf u}}
\def\vQ{{\bf Q}}
\def\vq{{\bf q}}
\def\vp{{\bf p}}
\def\vR{{\bf R}}
\def\vP{{\bf P}}
\def\vK{{\bf K}}
\newcommand{\bd}{\begin{equation}}
\newcommand{\ed}{\end{equation}}
\newcommand{\be}{\begin{equation}}
\newcommand{\ee}{\end{equation}}
\newcommand{\bt}{\begin{split}}
\newcommand{\et}{\end{split}}

\newcommand{\bn}{\begin{align}}
\newcommand{\en}{\end{align}}

\newcommand{\bea}{\begin{eqnarray}}
\newcommand{\eea}{\end{eqnarray}}
\newcommand{\ba}{\begin{array}}
\newcommand{\ea}{\end{array}}
\newcommand{\nn}{\nonumber}

\begin{document}
\title {Trion ground state, excited states and absorption spectrum using electron-exciton basis}

\author{Shiue-Yuan Shiau$^1$, Monique Combescot$^2$  and Yia-Chung Chang$^1$}
\email{yiachang@gate.sinica.edu.tw}
\affiliation{$^1$Research Center for Applied Sciences, Academia Sinica, Taipei, 115 Taiwan}
\affiliation{$^2$Institut des NanoSciences de Paris, Universit\'e Pierre et Marie Curie, CNRS, 4 place Jussieu, 75005 Paris}
\date{\today}


\begin{abstract}

We solve the Schr\"{o}dinger equation for two electrons plus one hole by writing it in the electron-exciton basis. The main advantage of this basis is to eliminate the exciton contribution from the trion energy in a natural way. The interacting electron-exciton system is treated using the recently developed composite boson many-body formalism which allows an exact handling of electron exchange. We numerically solve the resulting electron-exciton Schr\"{o}dinger equation, with the exciton levels restricted to the lowest $1s, 2s$ and $3s$ states, and we derive the trion ground state energy as a function of the electron-to-hole mass ratio. While our results are in reasonable agreement with those obtained through the best variational methods using free carrier basis, this electron-exciton basis is mostly suitable to easily reach the bound and unbound trion excited states. Through their wave functions, we here calculate the optical absorption spectrum in the presence of hot carriers for 2D quantum wells. We find large peaks located at the exciton levels, which are attributed to electron-exciton (unbound) scattering states, and small peaks identified with trion bound states.
\end{abstract}

\maketitle

\section{Introduction}
Trions, i.e., charged composite particles made of two electrons plus one hole ($X^-$), or two holes plus one electron ($X^+$), were first predicted in bulk semiconductors a few decades ago\cite{lampert1958}. These 3-body objects have well-known counterparts in atomic physics: charged hydrogen ion\cite{Bethe1957,Chandrasekhar1944,Frolov1998,Li2007} ${\rm H}^-$ or molecule ${\rm H}_2^+$. Although their binding energies are several orders of magnitude larger than for semiconductor trions, their physics are essentially the same provided that we use semiconductor effective masses, i.e., we neglect all complexities coming from the valence band by assuming one hole species only. \

 In the late 70's, trions with very small binding energy has been observed in bulk semiconductor samples such as Ge and Si (Refs.~\onlinecite{NaritaPRL1976,NaritaSSC1976}). However, it is in the 90's only that high-quality semiconductor quantum wells allowed precise studies of trion states using optical measurements\cite{Kheng1993,Finkelstein1995,Shields1995,Buhmann1995}. Indeed, the reduction of dimensionality is known to enhance all composite particle binding energies. Recent experiments in highly doped materials\cite{Brown1996,Kaur2000,Huard2000} have further investigated the interesting interaction which exists between trions and free carriers.

Although single trion eigenstates have up to now been approached through a free carrier basis made of two electrons plus one hole\cite{Wojs2007,Vukmirovc2008}---or two holes plus one electron---this basis is not the most appropriate one from a physical point of view. Indeed, the three free carriers of a trion interact through the strong Coulomb interaction which exists between elementary charges $\pm e$. This interaction has to be treated exactly in order to possibly find the poles associated with the trion bound states. It is physically clear that the trion ground state energy must lie slightly below the exciton energy, because the ``second'' electron is attracted not by an elementary charge but by the excitonic dipole. Yet, this weak attraction still has to be treated exactly in order to get the trion bound states. Thus, in the approach using free carriers, we face the challenging task of solving a 3-body Schr\"{o}dinger equation with a very high precision, since ultimately the trion binding energy we are interested in, only is a very small fraction of the total trion energy we get---the trion binding energy being the difference between the trion and the exciton ground state energies.\

The procedure we here propose avoids this difficulty. It uses a basis made of electron-exciton pairs $(e,X)$, the exciton states being analytically known, both in 2D and 3D. Instead of a free energy made of free carrier kinetic energies $\va^{(e)}_{\vk_e}+\va^{(e)}_{\vk^\prime_{e}}+\va^{(h)}_{\vk_h}$ with a minimum value equal to zero, the $(e,X)$ basis we here use already contains the exciton binding energy at zeroth order in the interaction, since the ``free energy'' then is $E_{X\vQ}+\va^{(e)}_{\vk^\prime_{e}}$. The construction of a trion Schr\"{o}dinger equation using electron-exciton pairs as a basis can be done by using the recently developed composite boson many-body theory\cite{moniqPhysReport} that can exactly handle carrier exchanges between exciton and free carrier as induced by the Pauli exclusion principle. \

The goal of this paper is not to compete with the variational approaches commonly used to derive the trion ground state energy. Thanks to rapid advance in computer technology, elaborate variational methods can now reach the ground state energy with an amazing precision of ten decimal digits or more\cite{Saavedra1998}. Rather, our purpose in proposing a new way to approach the trion problem is threefold:\

a) Variational methods are quite appropriate to derive the trion ground state energy with a numerical accuracy which surpasses experimental data. However, the coordinates commonly used\cite{Hylleraas1947} in these variational procedure completely mask the underlying physics of an electron interacting with an exciton through Coulomb processes and the effect of carrier exchanges on the structure. In contrast, the method presented here provides a physically clear picture of the many-body effects involved in a trion. Moreover, we can reach a reasonably good value of the trion ground state energy by numerically solving the trion Schr\"{o}dinger equation with just a few low-lying exciton levels kept in the electron-exciton basis, which makes this numerical resolution quite easy to perform.\

 b) The trion operator, defined in Eq.~(\ref{eq:wf_eX}), constructed on the electron-exciton basis we here use, is quite suitable for further study of interacting trions\cite{moniqEPJ2003,moniqSSC2003} because it integrates the trion into the general framework of interacting excitons, for which a powerful many-body formalism has been developed\cite{moniqPhysReport}. Solving the trion Schr\"{o}dinger equation in the electron-exciton basis, as addressed in the present work, provides useful grounds for future works on many-body effects involving trions.\

(c) Finally, the electron-exciton basis allows us to obtain the trion excited states as easily as the ground state, these excited states being difficult to reach within variational procedures.\

Using these ground and excited state wave functions makes it possible to compute the whole absorption spectrum of bound and unbound trions in doped semiconductors. We here concentrate on optical absorption in 2D quantum wells in the presence of hot carriers. For $X^{-}$ trion, i.e., two electrons and one hole with electron masses smaller than the hole mass, this spectrum shows one small peak below the ground state exciton energy, which is associated with the trion ground state, and right at the $1s,~2s$ and $3s$ exciton levels, larger peaks associated with unbound trions, i.e., electron-exciton scattering states. Through the carrier energy distribution, the absorption spectrum actually acts as a tool to probe the trion relative motion wave function in momentum space. The peak height is found to decrease with photon energy, due to the fact that the trion wave function amplitude decreases when the electron relative motion momentum increases (see Figs.~\ref{fig:WF_rpeta0} and \ref{fig:WF_r=0peta3}).  \

The paper is organized as follows: \

In Sec.\ref{sec:formalism_eeh}, we write the trion Schr\"{o}dinger equation using the two-electron-one-hole basis for completeness, since this basis is the standard one to approach trion. We pay particular attention to the parity condition for trions made of two electrons with same spin or opposite spins. \

In Sec.\ref{sec:formalism_eX}, we write the trion states in the electron-exciton basis and we show how the parity condition appears within this $(e,X)$ representation. We then derive the Schr\"{o}dinger equation fulfilled by the prefactor of the trion expansion in this $(e,X)$ basis. In the last part, we focus on trion made of opposite spin electrons in a singlet state to possibly reach the trion ground state and also the trion relative motion wave functions appearing in photon absorption spectrum.  \

In Sec.\ref{sec:result}, we first present numerical results on the binding energies of the trion ground and excited states for various electron-to-hole mass ratios. We then show the trion relative motion wave functions for bound and unbound states. Finally, we use these trion wave functions to calculate the optical absorption spectra for various hot carrier distributions. \

We then conclude.\

\section{$X^-$ Trion in terms of two electrons plus one hole $(e,e^\prime,h)$\label{sec:formalism_eeh}}

We consider two electrons with spins $s$ and $s^\prime$ and one hole with total angular momentum $m$. The electron spins can be $\pm 1/2$. In bulk samples, the hole angular momentum can be $m=(\pm 3/2, \pm 1/2)$, while in narrow quantum wells, it reduces to $m=\pm 3/2$ due to the heavy-light hole energy splitting induced by the well confinement. However, since the hole angular momentum does not play a role for a single $X^-$ trion if we neglect the exciton splitting due to electron-hole exchange, we will in the following ignore the hole angular momentum $m$ for simplicity. \

The basis commonly used to represent $X^-$ trion is made of three free carrier states, namely $a^\dag_{\vk_e; s}a^\dag_{\vk^\prime_e;s^\prime}b^\dag_{\vk_h}|v\ran$. By noting that the system Hamiltonian for two electrons $(e,e')$ plus one hole $(h)$ in first quantization,
\bea
H_{ee^\prime h}&=& \frac{\vp_e^2+\vp_{e^\prime}^2}{2m_e}+\frac{\vp_h^2}{2m_h}+\frac{e^2}{\epsilon_{sc}\left|\vr_e-\vr_{e^\prime}\right|}\nn\\
&&-\frac{e^2}{\epsilon_{sc}\left|\vr_e-\vr_h\right|}-\frac{e^2}{\epsilon_{sc}\left|\vr_{e^\prime}-\vr_h\right|}\label{eq:ham_eeh}
\eea
is invariant under the $(e\leftrightarrow e^\prime)$ exchange, where $\epsilon_{sc}$ is the semiconductor dielectric constant. We conclude that the orbital part of the trion wave function must be even or odd with respect to this exchange---the ground state wave function being even in order for this state to be nondegenerate. As a result, the trion eigenstates can be divided into two groups.  One group consists of states that have an even orbital part and odd spin part (due to the Pauli exclusion principle). Those are spin singlet $|S=0, S_z=0\ran$. The other group consists of states that have an odd orbital part and even spin part. Those are spin triplets $|S=1,S_z=(0,\pm 1)\ran$.\

 Consequently, the eigenstates of the Schr\"{o}dinger equation for one trion in the $(e,e^\prime,h)$ basis,
 \be
 (H_{ee^\prime h}-E^{(\mathcal{T},S)})|\Psi^{(\mathcal{T})}_{SS_z}\ran=0,\label{eq:ham_eeh02}
 \ee
 can be written as
\bea
\left|\Psi^{(\mathcal{T})}_{S=1,S_z=2s}\right\ran&=&\sum_{\vk_e\vk^\prime_e\vk_h}\psi^{(\mathcal{T},S=1)}_{\vk_e\vk^\prime_e\vk_h}a^\dag_{\vk_e; s}a^\dag_{\vk^\prime_e ;s}b^\dag_{\vk_h}|v\ran\label{eq:wf_samespin}
\eea
when the two electrons have the same spin $s$, and as
\bea
\left|\Psi^{(\mathcal{T})}_{S=(0,1),S_z=0}\right\ran&=&\sum_{\vk_e\vk^\prime_e\vk_h}\psi^{(\mathcal{T},S)}_{\vk_e\vk^\prime_e\vk_h}\left[a^\dag_{\vk_e; 1/2}a^\dag_{\vk^\prime_e; -1/2}\right.\nn\\
&&\left.-(-1)^Sa^\dag_{\vk_e; -1/2}a^\dag_{\vk^\prime_e; 1/2}\right]b^\dag_{\vk_h}|v\ran\label{eq:wf_diffspin}
\eea
when the two electron spins are opposite. \

Since $a^\dag_{\vk_e ;s}a^\dag_{\vk^\prime_e;s^\prime}=-a^\dag_{\vk^\prime_e; s^\prime}a^\dag_{\vk_e ;s}$, it is possible to replace the prefactors in Eq.~(\ref{eq:wf_samespin}) by $\left(\psi^{(\mathcal{T},S=1)}_{\vk_e\vk^\prime_e\vk_h}-\psi^{(\mathcal{T},S=1)}_{\vk^\prime_e\vk_e\vk_h}\right)/2$. For the same reason, we can rewrite the second operator in Eq.~(\ref{eq:wf_diffspin}) as the first one, the prefactors now reading as $(-1)^S\psi^{(\mathcal{T},S)}_{\vk^\prime_e\vk_e\vk_h}$. As a result, instead of writing the four trion states as in Eqs.~(\ref{eq:wf_samespin},\ref{eq:wf_diffspin}), we may as well write them in a more compact form as
\bea
\left|\Psi^{(\mathcal{T})}_{S,S_z=s+s^\prime}\right\ran&=&\sum_{\vk_e\vk^\prime_e\vk_h}\phi^{(\mathcal{T},S)}_{\vk_e\vk^\prime_e\vk_h}a^\dag_{\vk_e; s}a^\dag_{\vk^\prime_e ;s^\prime}b^\dag_{\vk_h}|v\ran,\label{eq:wf_sym}
\eea
where the prefactors in this expansion now read
\bea
\phi^{(\mathcal{T},S)}_{\vk_e\vk^\prime_e\vk_h}&=&\psi^{(\mathcal{T},S)}_{\vk_e\vk^\prime_e\vk_h}+(-1)^S\psi^{(\mathcal{T},S)}_{\vk^\prime_e\vk_e\vk_h}\label{fac:phi_varphi1}
\eea
within an irrelevant normalization factor. This new prefactor then fulfills the parity condition
\bea
\phi^{(\mathcal{T},S)}_{\vk_e\vk^\prime_e\vk_h}&=&(-1)^S\phi^{(\mathcal{T},S)}_{\vk^\prime_e\vk_e\vk_h},\label{eq:parity}
\eea
a requirement not necessary to enforce if we use the expansions (\ref{eq:wf_samespin},\ref{eq:wf_diffspin}) with prefactors $\psi^{(\mathcal{T},S)}_{\vk_e\vk^\prime_e\vk_h}$, since this parity condition is then fulfilled by the operators themselves.\

Due to the Schr\"{o}dinger equation (\ref{eq:ham_eeh02}), the prefactors of the trion state in Eq.~(\ref{eq:wf_sym}) must fulfill

\bea
0&=&\sum_{\vk_e\vk^\prime_e\vk_h}\Bigg\{\left[E_{\vk_e\vk^\prime_e\vk_h}-E^{(\mathcal{T},S)}\right]\phi^{(\mathcal{T},S)}_{\vk_e\vk^\prime_e\vk_h}\nn\\
&&+\sum_\vq V_\vq\left[\phi^{(\mathcal{T},S)}_{\vk_e+\vq,\vk^\prime_e-\vq,\vk_h}-\phi^{(\mathcal{T},S)}_{\vk_e+\vq,\vk^\prime_e,\vk_h-\vq}\right.\nn\\
&&\left.-\phi^{(\mathcal{T},S)}_{\vk_e,\vk^\prime_e+\vq,\vk_h-\vq}\right]\Bigg\}a^\dag_{\vk_e; s}a^\dag_{\vk^\prime_e; s^\prime}b^\dag_{\vk_h}|v\ran,\label{eq:Schrod_sym}
\eea
where the free part of the energy reads as $E_{\vk_e\vk^\prime_e\vk_h}=\va^{(e)}_{\vk_e}+\va^{(e)}_{\vk^\prime_e}+\va^{(h)}_{\vk_h}$. If we now project this equation onto $\lan v| b_{\vp_h}a_{\vp^\prime_e; s^\prime}a_{\vp_e; s}$ and use
\bd
\lan v| a_{\vp^\prime_e; s^\prime}a_{\vp_e; s}a^\dag_{\vk_e; s}a^\dag_{\vk^\prime_e; s^\prime}|v\ran=\delta_{\vp^\prime_e\vk^\prime_e}\delta_{\vp_e\vk_e}-\delta_{ss^\prime}\delta_{\vp^\prime_e\vk_e}\delta_{\vp_e\vk^\prime_e},\nn
\ed
we get one term only for $s^\prime=-s$, i.e., for $(S=(0,1),S_z=0)$ trions, namely
\bea
0&=&\left[E_{\vp_e\vp^\prime_e\vp_h}-E^{(\mathcal{T},S)}\right]\phi^{(\mathcal{T},S)}_{\vp_e\vp^\prime_e\vp_h}+\sum_\vq V_\vq\left[\phi^{(\mathcal{T},S)}_{\vp_e+\vq,\vp^\prime_e-\vq,\vp_h}\right.\nn\\
&&\left.-\phi^{(\mathcal{T},S)}_{\vp_e+\vq,\vp^\prime_e,\vp_h-\vq}-\phi^{(\mathcal{T},S)}_{\vp_e,\vp^\prime_e+\vq,\vp_h-\vq}\right].\label{eq:schrod_diffspin}
\eea
By contrast, for $s^\prime=s$, i.e., for $(S=1,S_z=2s)$ triplet trions, this projection yields two terms. These two terms, however, are equal, since $\phi^{(\mathcal{T},S=1)}_{\vk_e\vk^\prime_e\vk_h}=-\phi^{(\mathcal{T},S=1)}_{\vk^\prime_e\vk_e\vk_h}$ and $V_\vq=V_{-\vq}$, so that we end up with the same Schr\"{o}dinger equation (\ref{eq:schrod_diffspin}) for all four trion states. \

It is actually possible to recover this Schr\"{o}dinger equation starting from Eqs.~(\ref{eq:wf_samespin},\ref{eq:wf_diffspin}) in which the trion states are written in terms of $\psi^{(\mathcal{T},S)}_{\vk_e\vk^\prime_e\vk_h}$, instead of using Eq.~(\ref{eq:wf_sym}) with prefactors $\phi^{(\mathcal{T},S)}_{\vk_e\vk^\prime_e\vk_h}$. The derivation can be found in \ref{AppendixA}. \

The energies of trion states using the $(e,e^\prime,h)$ basis then follow from solving a 3-body Schr\"{o}dinger equation (\ref{eq:schrod_diffspin}) with the parity condition $\phi^{(\mathcal{T},S)}_{\vk_e\vk^\prime_e\vk_h}=(-1)^S\phi^{(\mathcal{T},S)}_{\vk^\prime_e\vk_e\vk_h}$. The trion ground state belongs to the set of even states, i.e., states with $S=0$, while the solutions with odd wave functions lead to triplet states, these three states being degenerate.

\section{Trion in terms of electron-exciton pairs $(e,X)$\label{sec:formalism_eX}}

We now turn to the trion representation in terms of electron-exciton pairs. Some useful results on excitons interacting with free electrons are briefly rederived in \ref{app:eqs}. References to equations in this Appendix start with the letter B.\

Since excitons form a complete basis for $(e,h)$ pairs, we can rewrite the trion states of Eq.~(\ref{eq:wf_sym}) as
\bea
\left|\Psi^{(\mathcal{T})}_{S,S_z=s+s^\prime}\right\ran&=&\sum_{\vk_e i}\phi^{(\mathcal{T},S)}_{\vk_e, i}a^\dag_{\vk_e; s}B^\dag_{i; s^\prime}|v\ran,\label{eq:wf_eX}
\eea
where Eq.~(\ref{app:ehpairs}) allows us to relate the prefactors of this $(e,X)$ expansion to the ones of the $(e,e^\prime,h)$ expansion in Eq.(\ref{eq:wf_sym}) through
\bea
\phi^{(\mathcal{T},S)}_{\vk_e, i}=\sum_{\vk^\prime_e\vk_h}\lan i|\vk^\prime_e\vk_h\ran\phi^{(\mathcal{T},S)}_{\vk_e\vk^\prime_e\vk_h}.
\eea
\

\begin{figure}[t]
\begin{center}
\includegraphics[trim=6cm 6cm 6cm 5cm,clip,width=2.8in] {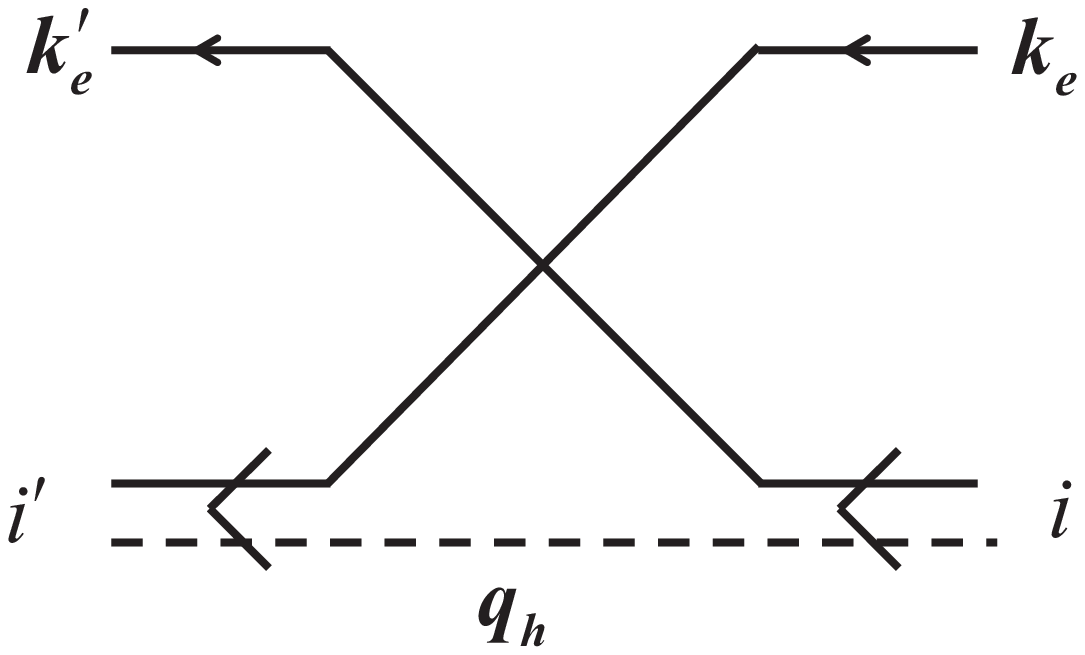}
\end{center}
\caption{{\small Pauli scattering $\lambda\big(^{~\vk_e^\prime~ \vk_e~}_{~
i^\prime\quad i~}\big)$ for electron exchange between a $\vk_e$ electron and a $i$ exciton, as given in Eq.~(\ref{app:func_lambda}). }}\label{fig:lambda}
\end{figure}

Before going further, let us note that the parity condition (\ref{eq:parity}) on $\phi^{(\mathcal{T},S)}_{\vk_e\vk^\prime_e\vk_h}$, resulting from the Pauli exclusion principle between the two electrons of the trion, imposes
\bea
&&\sum_{\vk_e i}\lambda\big(\begin{smallmatrix}
\vk_e^\prime& \vk_e\\
i^\prime& i
\end{smallmatrix}\big)\phi_{\vk_e, i}^{(\mathcal{T},S)}\nn\\
&&=\sum_{\vk_e i}\sum_{\vq_h}\sum_{\vp_e \vp_h}\lan i^\prime|\vk_e\vq_h\ran \lan \vq_h\vk^\prime_e|i\ran \lan i|\vp_e\vp_h\ran \phi_{\vk_e \vp_e\vp_h}^{(\mathcal{T},S)}\nn\\
&&=\sum_{\vk_e \vp_h}\phi_{\vk_e \vk^\prime_e\vp_h}^{(\mathcal{T},S)}\lan i^\prime|\vk_e\vp_h\ran=(-1)^S\sum_{\vk_e \vp_h}\phi_{\vk^\prime_e \vk_e\vp_h}^{(\mathcal{T},S)}\lan i^\prime|\vk_e\vp_h\ran\nn\\
&&=(-1)^S\phi_{\vk^\prime_e, i^\prime}^{(\mathcal{T},S)},
\eea
 where $\lambda\big(\begin{smallmatrix}
\vk_e^\prime& \vk_e\\
i^\prime& i
\end{smallmatrix}\big)$ is the Pauli scattering for electron exchange between a $\vk_e$ electron and a $i$ exciton, defined in Eq.~(\ref{app:func_lambda}) and shown in the diagram of Fig.~\ref{fig:lambda}. So, the parity condition (\ref{eq:parity}) reads in the $(e,X)$ subspace as
\bea
\phi_{\vk^\prime_e, i^\prime}^{(\mathcal{T},S)}&=&(-1)^S\sum_{\vk_e i}\lambda\big(\begin{smallmatrix}
\vk_e^\prime& \vk_e\\
i^\prime& i
\end{smallmatrix}\big)\phi_{\vk_e, i}^{(\mathcal{T},S)},\label{eq:parity02}
\eea
which clearly is more complicated than Eq.~(\ref{eq:parity}). We will show below how to overcome this complication.\

\begin{figure}[t]
\begin{center}
\includegraphics[trim=3cm 6cm 4cm 5cm,clip,width=3.4in] {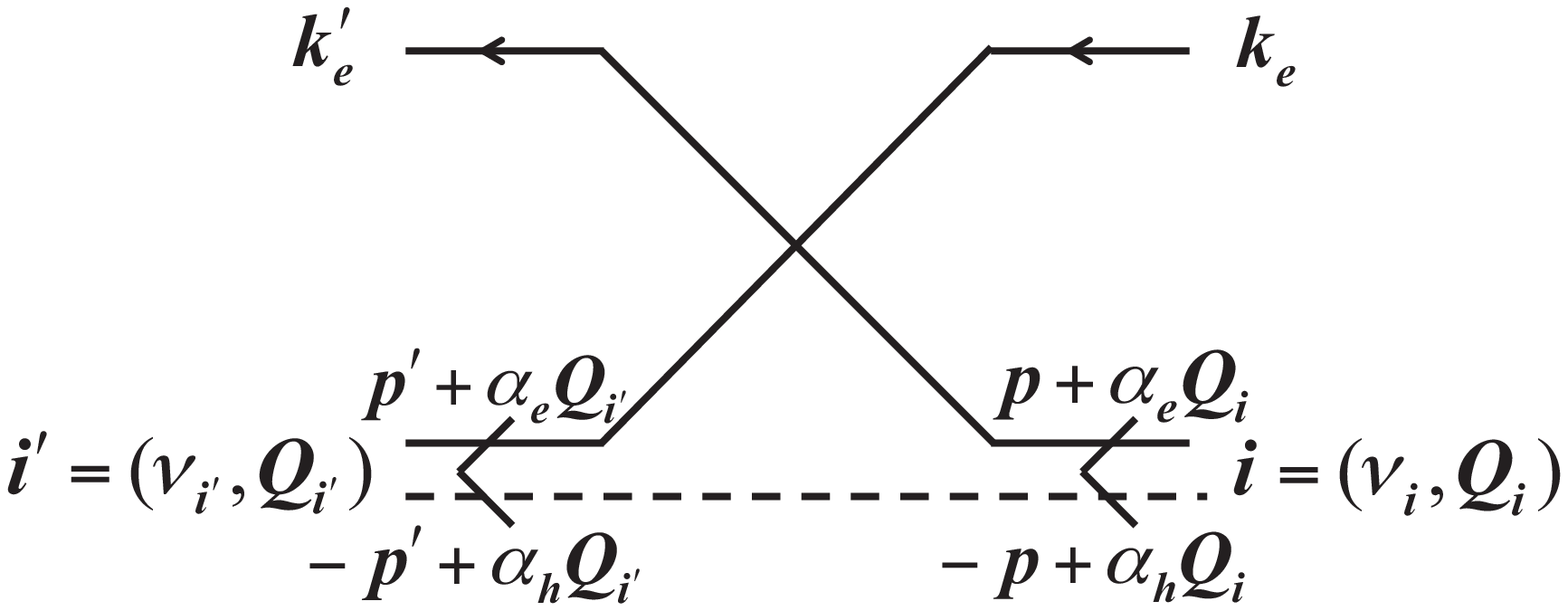}
\end{center}
\caption{{\small Pauli scattering $\lambda\big(^{~\vk_e^\prime~ \vk_e~}_{~
i^\prime\quad i~}\big)$ for electron exchange between a $\vk_e$ electron and a $i=(\nu_i,\vQ_i)$ exciton, as given in Eq.~(\ref{app:func_lambda2}). }}\label{fig:lambda2}
\end{figure}

\begin{figure}[t]
\begin{center}
\includegraphics[trim=4.5cm 6cm 3cm 6.5cm,clip,width=3.2in] {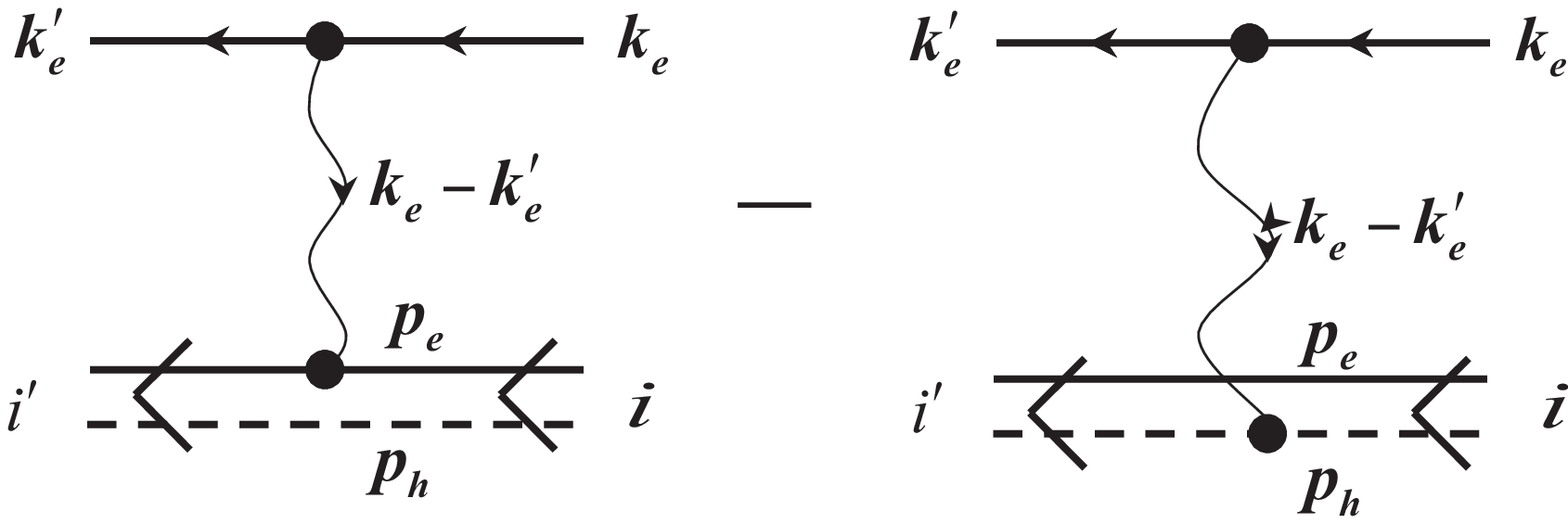}
\end{center}
\caption{{\small Direct Coulomb scattering $\xi\big(^{~\vk_e^\prime~ \vk_e~}_{~
i^\prime\quad i~}\big)$ between a $\vk_e$ electron and a $i$ exciton, as given in Eq.~(\ref{app:direcCoulomb00}).}}
\label{fig:Xi}
\end{figure}

For $H$ being the Hamiltonian of a system with arbitrary numbers of electrons and holes, we get using Eqs.~(\ref{app:commu_B}) and (\ref{app:commu_V})
\bea
\lefteqn{HB^\dag_{i;s^\prime}a^\dag_{\vk_e; s}|v\ran=\left(\left[H,B^\dag_{i; s^\prime}\right]+B^\dag_{i; s^\prime}H\right)a^\dag_{\vk_e; s}|v\ran}&&\nn\\
&&=E_{\vk_e, i}a^\dag_{\vk_e; s}B^\dag_{i; s^\prime}|v\ran+\sum_{\vk_e^\prime i^\prime} \xi\big(\begin{smallmatrix}
\vk_e^\prime& \vk_e\\
i^\prime& i
\end{smallmatrix}\big)a^\dag_{\vk^\prime_e; s}B^\dag_{i^\prime; s^\prime}|v\ran\label{eq:HBa}
\eea
where $E_{\vk_e, i}=\va_{\vk_e}^{(e)}+E_i^{(X)}$ is the free electron-exciton pair energy. The electron-exciton scattering $\xi\big(\begin{smallmatrix}
\vk_e^\prime& \vk_e\\
i^\prime& i
\end{smallmatrix}\big)$ associated with their direct Coulomb interactions is shown in the diagram of Fig.~\ref{fig:Xi}. \

From this equation, it is easy to show that the $|\Psi_{S, S_z=s+s^\prime}^{(\mathcal{T})}\ran$ state in Eq.~(\ref{eq:wf_eX}) is eigenstate of the Hamiltonian $H$ with energy $E^{(\mathcal{T},S)}$ provided that
\bea
0&=&\sum_{\vk_e i}\Bigg[\left(E_{\vk_e, i}-E^{(\mathcal{T},S)}\right)\phi_{\vk_e, i}^{(\mathcal{T},S)}\nn\\
&&+\sum_{\vk_e^\prime i^\prime} \xi\big(\begin{smallmatrix}
\vk_e& \vk^\prime_e\\
i& i^\prime
\end{smallmatrix}\big)\phi_{\vk^\prime_e, i^\prime}^{(\mathcal{T},S)}\Bigg]a^\dag_{\vk_e ;s}B^\dag_{i ;s^\prime}|v\ran.\label{eq:Schrod_eX01}
\eea

(i) Let us first consider $s=-s^\prime=1/2$. The projection of the above equation onto $\lan v|B_{j;-1/2}a_{\vp_e; 1/2}$ readily gives, using Eq.~(\ref{app:scalarprod_eX}),
\bd
0=\left(E_{\vp_e, j}-E^{(\mathcal{T},S)}\right)\phi_{\vp_e, j}^{(\mathcal{T},S)}+\sum_{\vk_e^\prime i^\prime}  \xi\big(\begin{smallmatrix}
\vp_e& \vk^\prime_e\\
j& i^\prime
\end{smallmatrix}\big)\phi_{\vk^\prime_e, i^\prime}^{(\mathcal{T},S)}.\label{eq:Schrod_eX02}
\ed
If instead, we project Eq.~(\ref{eq:Schrod_eX01}) onto $\lan v|B_{j; 1/2}a_{\vp_e; -1/2}$, we get a somewhat more complicated equation, namely,
\bea
0&=&\sum_{\vk_e i}\lambda\big(\begin{smallmatrix}
\vp_e& \vk_e\\
j& i
\end{smallmatrix}\big)\Bigg[\left(E_{\vk_e, i}-E^{(\mathcal{T},S)}\right)\phi_{\vk_e, i}^{(\mathcal{T},S)}\label{eq:Schrod_eX03}\\
&&+\sum_{\vk_e^\prime i^\prime}  \xi\big(\begin{smallmatrix}
\vk_e& \vk^\prime_e\\
i& i^\prime
\end{smallmatrix}\big)\phi_{\vk^\prime_e, i^\prime}^{(\mathcal{T},S)}\Bigg],\nn
\eea
which also reads as
\be
0=\sum_{\vk_e i}\Bigg[\lambda\big(\begin{smallmatrix}
\vp_e& \vk_e\\
j& i
\end{smallmatrix}\big)\left(E_{\vk_e, i}-E^{(\mathcal{T},S)}\right)+  \xi^{\rm in}\big(\begin{smallmatrix}
\vp_e& \vk_e\\
j& i
\end{smallmatrix}\big)\Bigg]\phi_{\vk_e, i}^{(\mathcal{T},S)},\label{eq:Schrod_eX030}
\ee
where $\xi^{\rm in}\big(\begin{smallmatrix}
\vp_e& \vk_e\\
j& i
\end{smallmatrix}\big)$ is the exchange-Coulomb scattering defined in Eq.~(\ref{eq:xi_in}), where ``in" refers to the fact that Coulomb interactions take place between the ``in'' electron-exciton pair $(\vk_e, i)$, as shown in the diagram of Fig.~\ref{fig:Xi_in}. It is actually possible to show that Eq.~(\ref{eq:Schrod_eX030}) follows from Eq.~(\ref{eq:Schrod_eX02}) by using the parity condition (\ref{eq:parity02}). Indeed, Eq.~(\ref{eq:parity02}) allows us to transform the RHS of Eq.~(\ref{eq:Schrod_eX02}) into
\be
 (-1)^S\sum_{\vk_e i}\Bigg[(E_{\vp_e, j}-E^{(\mathcal{T},S)})\lambda\big(\begin{smallmatrix}
\vp_e& \vk_e\\
j& i
\end{smallmatrix}\big)+ \xi^{\rm out}\big(\begin{smallmatrix}
\vp_e& \vk_e\\
j& i
\end{smallmatrix}\big)\Bigg]\phi_{\vk_e, i}^{(\mathcal{T},S)}\label{eq:Schrod_eX031}
\ee
where $\xi^{\rm out}\big(\begin{smallmatrix}
\vp_e& \vk_e\\
j& i
\end{smallmatrix}\big)$ is defined in Eq.~(\ref{eq:xi_out}). Using Eq.~(\ref{eq:rel_Pauliexchange}) for the link between $\xi^{\rm in}$ and $\xi^{\rm out}$, we then find that the $\phi_{\vk_e, i}^{(\mathcal{T},S)}$ prefactors in Eqs. (\ref{eq:Schrod_eX030}) and (\ref{eq:Schrod_eX031}) are indeed equal.\

(ii) We now consider Eq.~(\ref{eq:Schrod_eX01}) for $s=s^\prime$ and project it onto $\lan v|B_{j; s}a_{\vp_e; s}$. According to Eq.~(\ref{app:scalarprod_eX}), we get two terms. These two terms just correspond to the RHS of Eqs.~(\ref{eq:Schrod_eX02}) and (\ref{eq:Schrod_eX03}), so that we again end up with $\phi_{\vp_e, j}^{(\mathcal{T},S)}$ fulfilling Eq.~(\ref{eq:Schrod_eX02}) for $S_z=\pm1$. This is fully reasonable because Eq.~(\ref{eq:Schrod_eX02}) holds for triplet and singlet trions made of opposite spin electrons, while the three triplet states $(S=1,S_z=(0,\pm 1))$ are degenerate. So, they must obey the same Schr\"{o}dinger equation. \

As a result, the Schr\"{o}dinger equation for one trion in the electron-exciton basis reads for all trion states as
\bd
0=\left(E_{\vk_e, i}-E^{(\mathcal{T},S)}\right)\phi_{\vk_e, i}^{(\mathcal{T},S)}+\sum_{\vk_e^\prime i^\prime}  \xi\big(\begin{smallmatrix}
\vk_e& \vk^\prime_e\\
i& i^\prime
\end{smallmatrix}\big)\phi_{\vk^\prime_e, i^\prime}^{(\mathcal{T},S)},\label{eq:Schrod_eX04}
\ed
where the $\phi_{\vk_e, i}^{(\mathcal{T},S)}$ wave function in this basis is subject to the constraint imposed by the parity condition (\ref{eq:parity02}). The singlet and triplet trion states are given by Eq.~(\ref{eq:wf_eX}) with the prefactor $\phi^{(\mathcal{T},S)}_{\vk_e, i}$ fulfilling the Schr\"{o}dinger equation (\ref{eq:Schrod_eX04}). \

\begin{figure}[t]
\begin{center}
\includegraphics[trim=3cm 6cm 4cm 5cm,clip,width=3.2in] {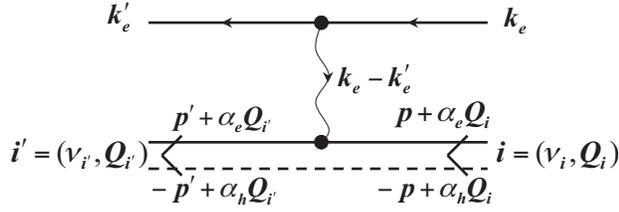}
\end{center}
\caption{{\small Part of the direct Coulomb scattering $\xi\big(^{~\vk_e^\prime~ \vk_e~}_{~
i^\prime\quad i~}\big)$ between a $\vk_e$ electron and a $i=(\nu_i,\vQ_i)$ exciton, as given in Eq.~(\ref{app:direcCoulomb01}), coming from electron-electron repulsion. $\xi\big(^{~\vk_e^\prime~ \vk_e~}_{~
i^\prime\quad i~}\big)$ also contains a negative contribution coming from electron-hole attraction, as seen from the second diagram of Fig.~\ref{fig:Xi}.}}
\label{fig:Xi2}
\end{figure}

The parity condition (\ref{eq:parity02}), however, is difficult to implement numerically. It is actually possible to include this condition into the Schr\"{o}dinger equation itself by setting
\bd
\phi^{(\mathcal{T},S)}_{\vk_e, i}=\varphi^{(\mathcal{T},S)}_{\vk_e, i}+(-1)^S\sum_{\vk_e^\prime i^\prime}\lambda\big(\begin{smallmatrix}
\vk_e& \vk_e^\prime\\
i& i^\prime
\end{smallmatrix}\big)\varphi_{\vk_e^\prime, i^\prime}^{(\mathcal{T},S)}\label{eq:wf_eX2};
\ed
so, the parity condition (\ref{eq:parity02}) on $\phi^{(\mathcal{T},S)}_{\vk_e, i}$ is automatically satisfied for arbitrary $\varphi^{(\mathcal{T},S)}_{\vk_e, i}$, which can then be regarded as a free function. By inserting Eq.~(\ref{eq:wf_eX2}) into the Schr\"{o}dinger equation (\ref{eq:Schrod_eX04}), we find the equation fulfilled by the $\varphi^{(\mathcal{T},S)}_{\vk_e, i}$ function as
\bea
0&=&\left(E_{\vk_e, i}-E^{(\mathcal{T},S)}\right)\Big[\varphi^{(\mathcal{T},S)}_{\vk_e, i}+(-1)^S\sum_{\vk_e^\prime i^\prime}\lambda\big(\begin{smallmatrix}
\vk_e& \vk_e^\prime\\
i& i^\prime
\end{smallmatrix}\big)\varphi_{\vk_e^\prime, i^\prime}^{(\mathcal{T},S)}\Big]\nn\\
&&+\sum_{\vk_e^\prime i^\prime} \Big[ \xi\big(\begin{smallmatrix}
\vk_e& \vk^\prime_e\\
i& i^\prime
\end{smallmatrix}\big)+ (-1)^S  \xi^{\rm out}\big(\begin{smallmatrix}
\vk_e& \vk^\prime_e\\
i& i^\prime
\end{smallmatrix}\big) \Big]\varphi_{\vk^\prime_e, i^\prime}^{(\mathcal{T},S)}.
\eea
Using Eq.~(\ref{eq:rel_Pauliexchange}), it is possible to rewrite this equation into a symmetrical form with respect to the $(\vk_e,i)$ and $(\vk_e^\prime,i^\prime)$ states, as
\bea
&&E_{\vk_e, i}\varphi^{(\mathcal{T},S)}_{\vk_e, i}+\sum_{\vk_e^\prime i^\prime} \Big[ \xi\big(\begin{smallmatrix}
\vk_e& \vk^\prime_e\\
i& i^\prime
\end{smallmatrix}\big)+ (-1)^S  \xi_{\rm ech}\big(\begin{smallmatrix}
\vk_e& \vk^\prime_e\\
i& i^\prime
\end{smallmatrix}\big) \Big]\varphi_{\vk^\prime_e, i^\prime}^{(\mathcal{T},S)}\nn\\
&&=E^{(\mathcal{T},S)}\Big[\varphi^{(\mathcal{T},S)}_{\vk_e, i}+(-1)^S\sum_{\vk_e^\prime i^\prime}\lambda\big(\begin{smallmatrix}
\vk_e& \vk_e^\prime\\
i& i^\prime
\end{smallmatrix}\big)\varphi_{\vk_e^\prime, i^\prime}^{(\mathcal{T},S)}\Big],\label{eq:Sing_tscr}
\eea
where the relevant set of exchange processes now take a symmetrical form between the ``in'' state $(\vk_e^\prime, i^\prime)$ and ``out'' state $(\vk_e, i)$, namely
\bea
\xi_{\rm ech}\big(\begin{smallmatrix}
\vk_e& \vk^\prime_e\\
i& i^\prime
\end{smallmatrix}\big)&=&\frac{1}{2}\Big[ \xi^{\rm in}\big(\begin{smallmatrix}
\vk_e& \vk^\prime_e\\
i& i^\prime
\end{smallmatrix}\big)+ \xi^{\rm out}\big(\begin{smallmatrix}
\vk_e& \vk^\prime_e\\
i& i^\prime
\end{smallmatrix}\big)\nn\\
&&+\big(E_{\vk_e, i}+E_{\vk^\prime_e, i^\prime}\big)\lambda\big(\begin{smallmatrix}
\vk_e& \vk_e^\prime\\
i& i^\prime
\end{smallmatrix}\big) \Big].
\eea

\begin{figure}[t]
\begin{center}
\includegraphics[trim=4cm 8cm 3.5cm 6cm,clip,width=3.4in] {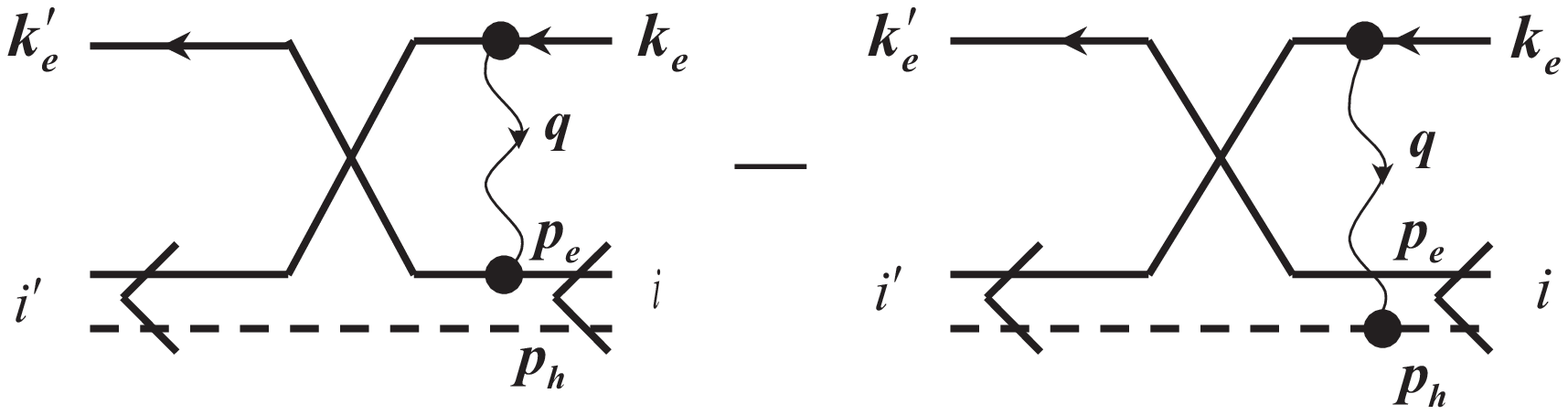}
\end{center}
\caption{{\small ``In'' exchange-Coulomb scattering $\xi^{\rm in}\big(^{~\vk_e^\prime~ \vk_e~}_{~
i^\prime\quad i~}\big)$ between a $\vk_e$ electron and a $i$ exciton, as given in Eq.~(\ref{app:inexCoulomb01}).}}
\label{fig:Xi_in}
\end{figure}

\begin{figure}[t]
\begin{center}
\includegraphics[trim=3.5cm 7cm 5.5cm 4cm,clip,width=3.4in] {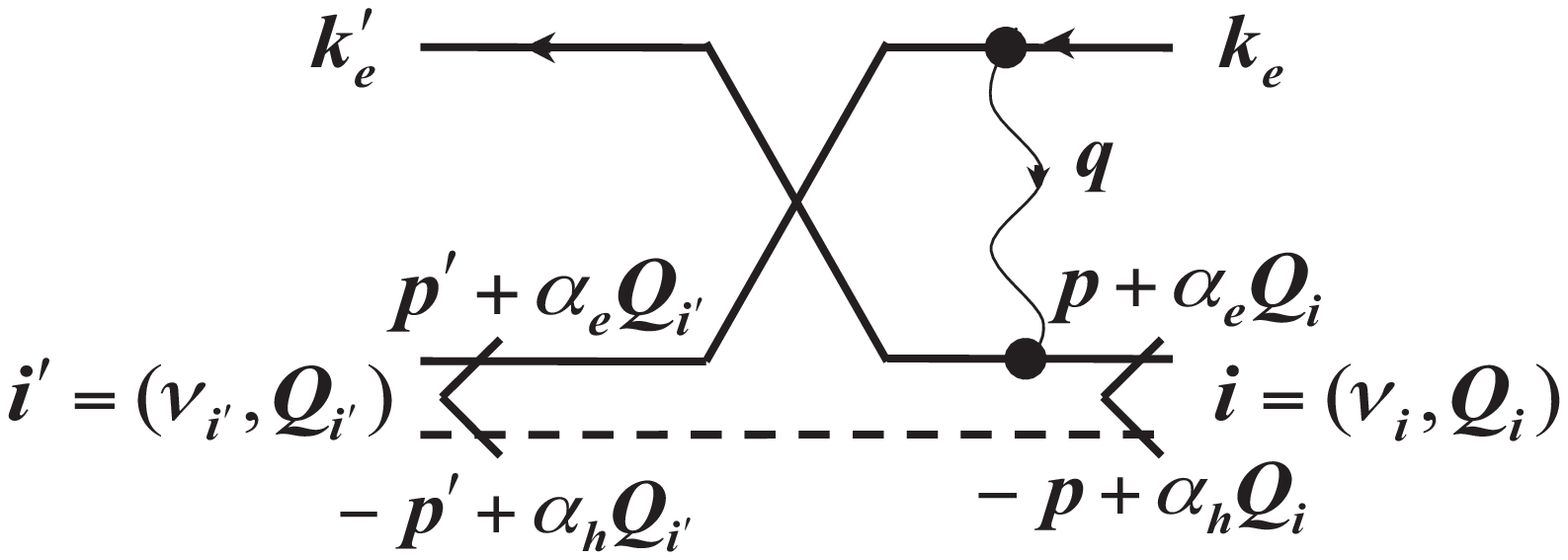}
\end{center}
\caption{{\small Part of the ``in'' exchange-Coulomb scattering $\xi^{\rm in}\big(^{~\vk_e^\prime~ \vk_e~}_{~
i^\prime\quad i~}\big)$ between a $\vk_e$ electron and a $i=(\nu_i,\vQ_i)$ exciton, as given in Eq.~(\ref{app:inexCoulomb02}), coming from electron-electron repulsion.}}
\label{fig:Xi_in2}
\end{figure}

We note that this effective exchange scattering not only contains the ``in'' and ``out'' Coulomb-exchange scatterings, but also the Pauli scattering multiplied by the sum of ``in'' and ``out'' electron-exciton pair energies in order to end with an energy-like quantity.\

From now on, we only consider trions with zero center-of-mass momentum. These are constructed on zero-momentum electron-exciton pairs, $\vk$ and $i=(\nu,-\vk)$. Moreover, we concentrate on singlet trion states, $S=0$, which are the relevant states for the determination of both the trion ground state and the photoabsorption spectrum in the presence of a doped electron gas. So, we will not quote the $S$ index anymore to simplify the notation.\

 Equation (\ref{eq:Sing_tscr}) can be formally written into a matrix form as
\begin{equation}
(\hat{E}+\hat{\xi}_{\rm eff})\varphi^{(\mathcal{T})}=E^{(\mathcal{T})}(\hat{1}+\hat{\lambda})\varphi^{(\mathcal{T})}\label{eq:schroeq_mf},
\end{equation}
where $E^{(\mathcal{T})}$ is the trion energy we want to determine. For zero center-of-mass momentum pairs, as considered from now on, the components of the $\varphi^{(\mathcal{T})}$ vector are $\varphi^{(\mathcal{T})}(\vk,\nu)\equiv\varphi^{(\mathcal{T})}_{\vk,i=(\nu,-\vk)}$. The identity matrix is denoted as $\hat 1$ while $\hat{E}$ is a diagonal matrix with components $\lan \vk\nu|\hat{E}|\vk^\prime \nu^\prime\ran=E(\vk,\nu)\delta_{\vk\vk^\prime}\delta_{\nu\nu^\prime}$, the free electron-exciton pair energy being $E(\vk,\nu)=E_{\vk,(\nu,-\vk)}=\va_{\nu}+\vk^2/2\mu_{eX}$ where $\mu_{eX}^{-1}=m_e^{-1}+(m_e+m_h)^{-1}$ is the inverse of the electron-exciton pair reduced mass. $\hat{\xi}_{\rm eff}$ is an energy-like matrix which includes Coulomb and exchange processes within the electron-exciton pair. Its components are
\bd
\lan \vk\nu|\hat{\xi}_{\rm eff}|\vk^\prime \nu^\prime\ran= \Big[\xi\big(\begin{smallmatrix}
\vk & \vk^\prime\\
(\nu,-\vk)& (\nu^\prime,-\vk^\prime)
\end{smallmatrix}\big)+ \xi_{\rm ech}\big(\begin{smallmatrix}
\vk & \vk^\prime\\
(\nu,-\vk)& (\nu^\prime,-\vk^\prime)
\end{smallmatrix}\big)\Big].
\ed
Finally, $\hat{\lambda}$ is a dimensionless matrix which originates from the fact that the hole in the electron-hole pair forming the exciton can be associated with any of the two electrons. Its components are $\lan \vk\nu|\hat\lambda|\vk^\prime \nu^\prime\ran= \lambda\big(\begin{smallmatrix}
\vk & \vk^\prime\\
(\nu,-\vk)& (\nu^\prime,-\vk^\prime)
\end{smallmatrix}\big)$.\

Figures 1-4 show the various scatterings appearing between an electron and an exciton (for their derivations, see \ref{app:eqs}). For zero-momentum electron-exciton pairs, they take a compact form: The electron exchange reduces to
\bd
\lambda\big(\begin{smallmatrix}
\vk& \vk^\prime\\
(\nu,-\vk)& (\nu^\prime,-\vk^\prime)
\end{smallmatrix}\big)=\lan \nu|\vk^\prime+\alpha_e\vk\ran\lan \vk+\alpha_e\vk^\prime|\nu^\prime\ran,\label{lambdaforK=0}
\ed
while the direct Coulomb scattering between an electron and an exciton simply reads as
\bea
\lefteqn{\xi\big(\begin{smallmatrix}
\vk& \vk^\prime\\
(\nu,-\vk)& (\nu^\prime,-\vk^\prime)\end{smallmatrix}\big)}\\
&=&V_{\vk-\vk^\prime}\sum_{{\vp}}\Big[\lan \nu|{\vp}-\alpha_h\vk \ran \lan {\vp}-\alpha_h\vk^\prime|\nu^\prime\ran-(\alpha_h\rightarrow-\alpha_e)\Big].\nn
\eea
When Coulomb scattering is mixed with electron exchange, the resulting exchange-Coulomb scattering is given, for zero-momentum pairs, by
\bea
\lefteqn{\xi^{\rm in}\big(\begin{smallmatrix}
\vk& \vk^\prime\\
(\nu,-\vk)& (\nu^\prime,-\vk^\prime)\end{smallmatrix}\big)
=\Big[\xi^{\rm out}\big(\begin{smallmatrix}
\vk^\prime& \vk\\
(\nu^\prime,-\vk^\prime)& (\nu,-\vk)\end{smallmatrix}\big)\Big]^*}\nn\\
&=&\sum_{\vp} V_{\vk^\prime+\alpha_e\vk-\vp}\lan\nu|\vp\ran\nn\\
&&\times\Big[\lan \vp+\alpha_h(\vk-\vk^\prime)|\nu^\prime\ran-\lan \vk+\alpha_e\vk^\prime|\nu^\prime\ran\Big],
\eea
where $V_\vq=2\pi e^2/\epsilon_{sc}L^2q$ in 2D and $V_\vq=4\pi e^2/\epsilon_{sc}L^3q^2$ in 3D.

It is actually possible to cast Eq.~(\ref{eq:schroeq_mf}) into a Schr\"{o}dinger-like equation for $\varphi^{(\mathcal{T})}$, by rewriting it as
\begin{equation}
E^{(\mathcal{T})}\varphi^{(\mathcal{T})}=(\hat 1+\hat{\lambda})^{-1}(\hat{E}+\hat{\xi}_{\rm eff})\varphi^{(\mathcal{T})}\equiv\hat{H}_{\rm eff}\varphi^{(\mathcal{T})}.\label{eq:schroeq_mf1}
\end{equation}
The operator $\hat{H}_{\rm eff}$ then appears as the effective Hamiltonian for trion written in the electron-exciton basis. We see that electron exchange within this pair appears not only through the energy-like exchange scatterings $(\hat\xi^{\rm in},\hat\xi^{\rm out})$ and $(E\hat\lambda)$ included in $\hat\xi_{\rm ech}$, but also in a more subtle way through the $(\hat 1+\hat{\lambda})^{-1}$ operator appearing in front of the more naive part of the Hamiltonian, namely $(\hat{E}+\hat{\xi}_{\rm eff})$.

\begin{figure}[t]
\begin{center}
\epsfig{figure=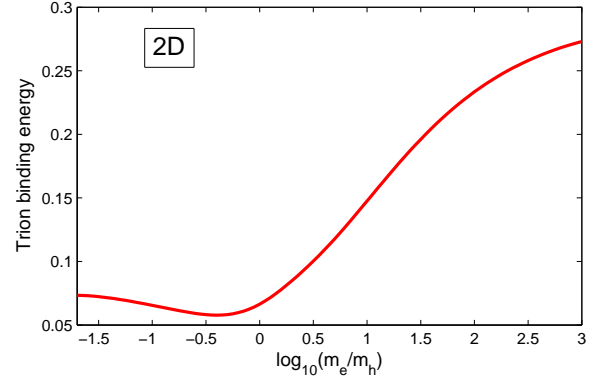,clip=,width=3.4 in}
\end{center}
\caption{(Color online) Ground state trion binding energy in unit of the 2D exciton binding energy $R_X^{(2D)}=4R_X^{(3D)}$, as a function of the electron-to-hole mass ratio. A minimum is found for $m_e$ slightly lighter than $m_h$.}
\label{fig:result_bindingEgy2D}
\end{figure}

\begin{figure}[t]
\begin{center}
\epsfig{figure=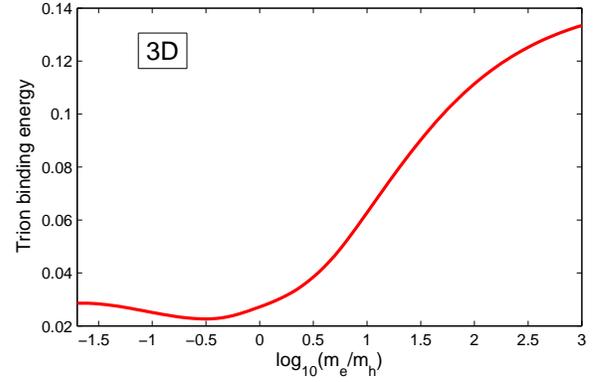,clip=,width=3.4 in}
\end{center}
\caption{(Color online) Same as Fig.~\ref{fig:result_bindingEgy2D} for 3D trion, the binding energy unit being $R_X^{(3D)}$. The behavior is qualitatively similar, but the binding energies are significantly smaller, which makes the experimental observation of 3D trions more difficult.}
\label{fig:result_bindingEgy3D}
\end{figure}

\section{Results and Discussions\label{sec:result}}

The last step is to numerically solve the Schr\"{o}dinger equation (\ref{eq:schroeq_mf}) or equivalently Eq.~(\ref{eq:schroeq_mf1}). To do so, we restrict the exciton basis $\lan \vp|\nu\ran$ to the three low-lying $s$-like states. The inclusion of $p$- or $d$-like exciton wave functions requires a very careful treatment of the angular part of the electron-exciton pair wave functions in the calculation of the $\hat\lambda$ and $\hat\xi$ scatterings, which is beyond the scope of the present work. This extension will be considered elsewhere. The 2D normalized wave functions $\lan \vp|\nu\ran$ used in this electron-exciton basis are\cite{Yang91}
\bea
\lan \vp|1s\ran&=& \left(\frac{a_X}{L}\right)\frac{\sqrt{2\pi}}{(1+a_X^2\vp^2/4)^{3/2}},\nn\\
\lan \vp|2s\ran&=& \left(\frac{a_X}{L}\right)\frac{3\sqrt{6\pi}(9a_X^2\vp^2/4-1)}{(1+9a_X^2\vp^2/4)^{5/2}},\\
\lan \vp|3s\ran&=& \left(\frac{a_X}{L}\right)\frac{5\sqrt{10\pi}(1-25a_X^2\vp^2+625a_X^4\vp^4/16)}{(1+25a_X^2\vp^2/4)^{7/2}},\nn
\eea
while for 3D they read as
\bea
\lan \vp|1s\ran&=& \left(\frac{a_X}{L}\right)^{3/2}\frac{8\sqrt{\pi}}{(1+a_X^2\vp^2)^2},\nn\\
\lan \vp|2s\ran&=& \left(\frac{a_X}{L}\right)^{3/2}\frac{32\sqrt{2\pi}(4a_X^2\vp^2-1)}{(1+4a_X^2\vp^2)^3},\\
\lan \vp|3s\ran&=& \left(\frac{a_X}{L}\right)^{3/2}\frac{72\sqrt{3\pi}(1-30a_X^2\vp^2+81a_X^4\vp^4)}{(1+9a_X^2\vp^2)^4}.\nn
\eea
$L$ is the sample size, $a_X$ is the 3D exciton Bohr radius defined as $a_X=\hbar^2\epsilon_{sc}/\mu_Xe^2$, with $\mu_X^{-1}=m_e^{-1}+m_h^{-1}$ being the inverse of the exciton reduced mass and $\epsilon_{sc}$ being the semiconductor dielectric constant, one order of magnitude larger in semiconductor samples than in vacuum. \

The Schr\"{o}dinger equation (\ref{eq:schroeq_mf1}), seen as a matrix spanned by the $\vk$ momentum, can be solved as a generalized eigenvalue problem. To this end, we sampled the $k=|\vk|$ value with $100$ mesh points, according to $k_i=u_i^3$ where the $u_i$'s are equally distributed, thereby allowing for more sampling in the small $k$ region. However, for $m_e/m_h=1/50$ and $1/100$ in 3D, we have taken $220$ and $350$ mesh points in order for the $\lambda$ matrices to unambiguously stay positive definite. We also took an upper cutoff $k_{\rm max}$ of $10$ in 3D but $20$ in 2D (in unit of $a_X$), since the 2D wave functions have a larger radial extension in {\bf k}-sspace. \

\subsection{Trion binding energies for ground and excited states}

Figures~\ref{fig:result_bindingEgy2D} and \ref{fig:result_bindingEgy3D} show the binding energies of the 2D and 3D ground state trions as a function of the electron-to-hole mass ratio $m_e/m_h$. These two sets of results are expressed in terms of their 2D and 3D effective Rydbergs, namely $R_X^{(2D)}$ and $R_X^{(3D)}$, with $R_X^{(2D)}=4R_X^{(3D)}$ and $R_X^{(3D)} = (\tilde \mu_X/\epsilon_{sc}^2) 13.6$~eV, the ratio of exciton reduced mass to free electron mass $m_0$ being $\tilde \mu_X=m_em_h/m_0(m_e+m_h)$. Although the qualitative behaviors of the 2D and 3D curves are very similar, it is worth noting that the absolute values of the 2D binding energies are significantly larger than in 3D. This is mainly due to the fact that the Coulomb interaction $V_\vq$ decreases more slowly, as $1/q$ in 2D instead of $1/q^2$ in 3D. For $m_e/m_h\lesssim 1$, the trion binding energy stays essentially constant as a function of mass ratio, in agreement with experimental data: in most bulk semiconductor samples like GaAs and InAs, with an effective electron mass smaller than the effective hole mass, the $X^-$ trion binding energy does not significantly depend on the mass ratio. This is also true for 2D.\

\begin{figure}[t]
\begin{center}
\epsfig{figure=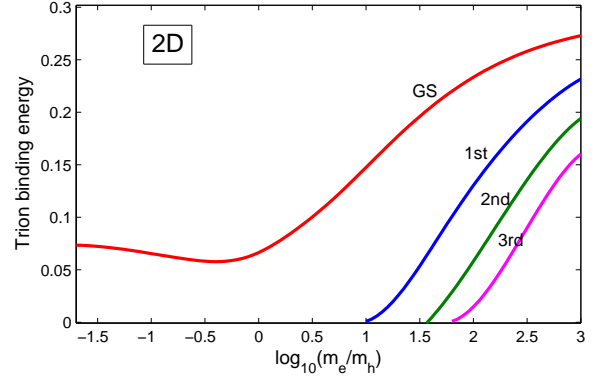,clip=,width=3.4 in}
\end{center}
\caption{(Color online) Bound state binding energies of the 2D trion ground state(GS) and the first 3 excited states in $R_X^{(2D)}$ unit as a function of the electron-to-hole mass ratio.}
\label{fig:result_DOS2D}
\end{figure}

\begin{figure}[t]
\begin{center}
\epsfig{figure=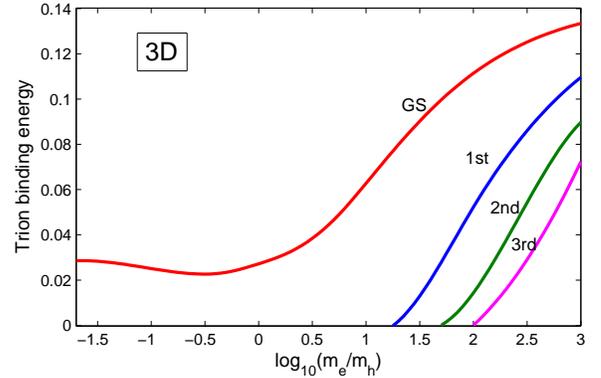,clip=,width=3.4 in}
\end{center}
\caption{(Color online) Same as Fig.~\ref{fig:result_DOS2D} for 3D, the energy unit being $R_X^{(3D)}$. }
\label{fig:result_DOS3D}
\end{figure}

In contrast, when $m_e/m_h$ increases above $1$, which mimics the change from $X^{-}$ to $X^{+}$ trions, with two carriers then having a heavy mass, we see that the binding energy first increases essentially logarithmically and then saturates for large mass ratios: when $m_e/m_h\gg 1$, the trion behaves as an hydrogen molecule-ion ${\rm H}_2^+$, with a strong binding energy known to be $0.412~R_X^{(2D)}$ in 2D (Ref.~\onlinecite{Zhu1990}) and $0.194~R_X^{(3D)}$ in 3D (Ref.~\onlinecite{Li2007}). \

The binding energies of the 2D and 3D trion ground states we obtain as a function of mass ratio, are qualitatively similar to those obtained from variational methods\cite{Stebe1989,Usukura1999,Thilagam1997,Sergeev2001}. However, the values we find only account for $60\%-70\%$ of the most accurate variational results when $m_e/m_h> 1$, and $50\%-60\%$ only when $m_e/m_h\lesssim 1$. This suggests that the polarization of the exciton induced by the second electron, which is not accounted for here because our numerical calculation excludes $p$-like, $d$-like, and higher angular momentum exciton wave functions, plays a significant role in the binding, and this is more so when two carriers have a light mass. For GaAs, with effective electron mass $m_e=0.067m_0$, effective hole mass $m_h=0.34m_0$, and dielectric constant $\epsilon_{sc}=12.5$, Fig.~\ref{fig:result_bindingEgy2D} gives a 2D trion binding energy equal to $0.06~R_X^{(2D)}=1.2~$meV, which is about a factor of 2 smaller than the best variational result\cite{Stebe1989,Thilagam1997,Sergeev2001}.
\

In spite of this discrepancy, one important advantage of the present method compared to usual variational approaches is that it allows to reach trion excited states as easily as the ground state, while the excited states are not easily obtained by usual variational methods. Figures~\ref{fig:result_DOS2D} and \ref{fig:result_DOS3D} show the various excited bound states of 2D and 3D trions as a function of the mass ratio $m_e/m_h$. For $m_e/m_h\leq 1$, we find one bound state only, which is the ground state, in agreement with known results for $X^-$ trions, with a hole mass larger than the electron mass, both in 3D (Ref.~\onlinecite{Hill1977}) and in 2D (Ref.~\onlinecite{Larsen1992}). By contrast, the number of trion bound states increases with $m_e/m_h$ when this ratio gets larger than 1. A second bound state emerges for $m_e/m_h\simeq 16$ in 3D and $m_e/m_h\simeq 10$ in 2D. And so on... These additional bound states mainly come from the vibrational motion of the two heavy carriers. The trion then is similar to a hydrogen molecule-ion ${\rm H}_2^+$ or its deuterium isotopes ${\rm HD}^+$ and ${\rm D}_2^+$. Spectroscopic studies of their vibrational and rotational levels have been a central issue in molecular physics because they represent the simplest molecules ever theoretically\cite{Stanley1960,Beckel1970,Bishop1973,Carrington1989} and experimentally\cite{Wing1976,Koelemeij2007,Critchley2001} studied. For very large mass ratio $m_e/m_h=1000$ close to the molecular limit, we find $15$ and $12$ bound states in 2D and 3D systems, respectively. \

\begin{figure}[t]
 \centering \subfigure[] {\ \label{fig:11a}
\includegraphics[width=3.4 in]{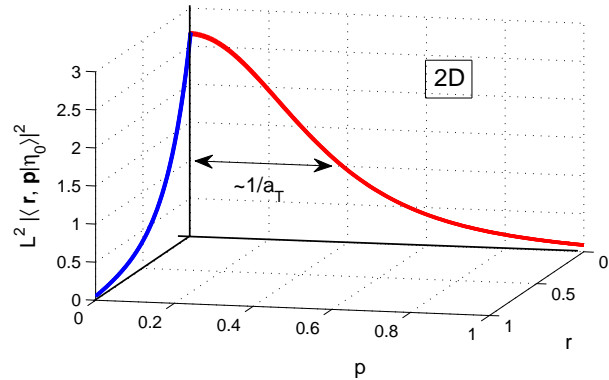} }\\
 \centering \subfigure[] {\ \label{fig:11b}
\includegraphics[width=3.4 in]{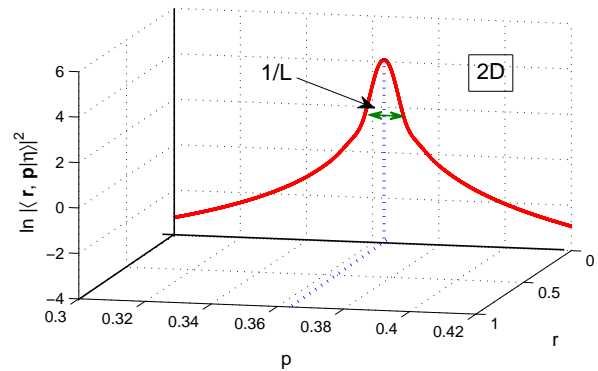}}\\
\caption{(Color online) $L^2|\langle \vr, \vp|\eta\rangle|^2$ for the trion bound ground state $\eta=\eta_0$ (figure a)
and $|\langle \vr, \vp|\eta\rangle|^2$ for an unbound state in a semilog plot (figure b). The mass ratio is taken equal to $m_h/m_e=5$. The units for relative motion momentum $p$ and exciton radial coordinate $r$ are $a_X$ and $a_X^{-1}$, respectively.}
\label{fig:WF_rpeta0}
\end{figure}

\subsection{Trion wave function}

By solving the Schr\"{o}dinger equation (\ref{eq:schroeq_mf1}), we can also obtain the $\varphi^{(\mathcal{T},S=0)}_{\vk,i=(\nu,-\vk)}$ function from which the trion wave function $\phi^{(\mathcal{T}, S=0)}_{\vk,i=(\nu,-\vk)}$ with proper symmetry property in the electron-exciton basis, can be reached through Eq.~(\ref{eq:wf_eX2}). Indeed, since Pauli scatterings $\lambda\big(\begin{smallmatrix}
\vk_e& \vk_e^\prime\\
i& i^\prime
\end{smallmatrix}\big)$ conserve momentum and spin, the $\varphi^{(\mathcal{T},S=0)}_{\vk_e^\prime,i^\prime}$ functions which are required in the sum of Eq.~(\ref{eq:wf_eX2}), also correspond to $\varphi^{(\mathcal{T}, S=0)}_{\vk_e^\prime,i^\prime=(\nu^\prime,-\vk_e^\prime)}$ functions.\

To go further, let us first relate the $\phi^{(\mathcal{T}, S=0)}_{\vk_e,i=(\nu,-\vk_e)}$ function, which we can derive through solving the Schr\"{o}dinger equation (\ref{eq:schroeq_mf1}), to the trion relative motion wave function used in previous works on trion. In Ref.~\onlinecite{MoniqueEPJ2004}, we have shown (see Eq.~(3.16) in this reference) that the wave function of a trion with center-of-mass momentum $\vK$, relative motion index $\eta$ and spin $S=(0,1)$ splits as
\be
\langle \vr_e,\vr_{e^\prime}, \vr_h|\vK, \eta ,S\rangle=\langle \vR_T|\vK\rangle\langle \vr, \vu|\eta, S\rangle,
\ee
where $\vR_T=(m_e\vr_e+m_e\vr_{e^\prime}+m_h \vr_h)/(2m_e+m_h)$ is the center-of-mass coordinate, $\vr=\vr_e-\vr_h$ is the distance between one electron and the hole, while $\vu=\vr_{e^\prime}-(m_e\vr_e+m_h \vr_h)/(m_e+m_h)$ is the distance between the other electron and the center-of-mass coordinate of the electron-hole pair. From $\langle \vr, \vu|\eta, S\rangle$, we can construct $\langle \nu, \vp|\eta, S\rangle$ through a double Fourier transform (see Eq.~(3.18) of Ref. \onlinecite{MoniqueEPJ2004})
\be
\langle \nu, \vp|\eta, S\rangle=\int d\vr d\vu \langle \nu|\vr \rangle\langle \vp|\vu \rangle \langle \vr, \vu|\eta, S\rangle.
\ee
The parity condition $\langle \vr_e,\vr_{e^\prime}, \vr_h|\vK ,\eta, S\rangle=(-1)^S\langle \vr_{e^\prime},\vr_e, \vr_h|\vK, \eta, S\rangle$ on the $(\vr_e,\vr_{e^\prime})$ coordinates then leads to
\be
\langle \nu, \vp|\eta, S\rangle=(-1)^S\sum_{\nu^\prime,\vp^\prime}\lan \nu|\vp^\prime+\alpha_e\vp\ran\lan \vp+\alpha_e\vp^\prime|\nu^\prime\ran\langle \nu^\prime, \vp^\prime|\eta, S\rangle
\ee
which exactly is the parity condition for $\phi^{(\mathcal{T}, S)}_{\vk_e,i}$ given in Eq.~(\ref{eq:parity02}) when $i=(\nu,-\vk_e)$, i.e., when the center-of-mass momentum of the electron-exciton pair is zero, because the Pauli scattering then reduces to Eq.~(\ref{lambdaforK=0}).\

If we now consider the creation operator for the $(\vK,\eta,S)$ trion made of opposite spin electrons, we find that it reads in terms of electron-exciton pairs as (see Eq.~(3.44) of Ref. \onlinecite{MoniqueEPJ2004})
\be
T^\dag_{\vK\eta S,S_z=0}=\sum_{\nu, \vp}\langle \nu, \vp|\eta, S\rangle a^\dag_{\vp+\beta_e\vK,1/2}B^\dag_{\nu,-\vp+\beta_X\vK,-1/2}.
\ee
So, in view of Eq.~(\ref{eq:wf_eX}), we are led to conclude that $\langle \nu, \vp|\eta, S\rangle$ must be identified with
\be
\langle \nu ,\vp|\eta, S\rangle=\phi^{(\mathcal{T}, S)}_{\vp,i=(\nu,-\vp)}
\ee
with $\mathcal{T}=(\vK=0,\eta)$.\

\begin{figure}[t]
 \begin{center}
\epsfig{figure=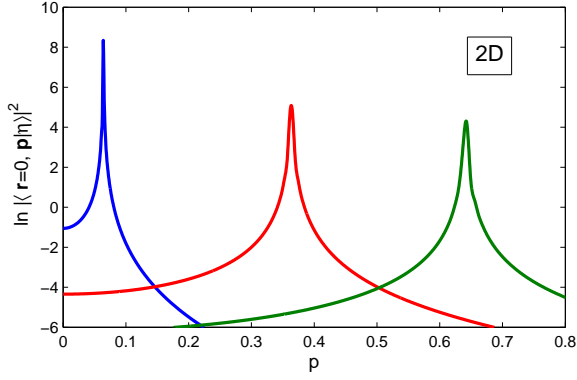,clip=,width=3.4 in}
\end{center}
\caption{(Color online) $|\langle \vr=0, \vp|\eta\rangle|^2 $ for three unbound trion states in a semilog plot, the mass ratio being $m_h/m_e=5$ and the momentum $p$ unit being $a_X$.}
\label{fig:WF_r=0peta3}
\end{figure}

Actually, the relative motion wave functions that have physical relevance is not $\langle \nu, \vp|\eta, S\rangle$ but $\langle \vr, \vu|\eta ,S\rangle$ or $\langle \vr, \vp|\eta ,S\rangle$. For bound state trion, $\langle \vr, \vu|\eta, S\rangle$ has an extension of the order of the exciton size $a_X$  along $\vr$ and an extension of the order of the trion size $a_T$ along $\vu$, while this latter extension goes up to the sample size $L$ for unbound trions. Through a Fourier transform, we can reach
\be
\langle \vr, \vp|\eta, S\rangle=\int  d\vu \langle \vp|\vu \rangle \langle \vr, \vu|\eta, S\rangle,\label{WF:rp_ru}
\ee
which has an extension along $\vp$ of the order of $1/a_T$ for bound state (see Fig.~\ref{fig:11a}) and of the order of $1/L$ for unbound states (see Figs.~\ref{fig:11b} and \ref{fig:WF_r=0peta3}).\

Due to dimensional arguments, the normalized $|\langle \vr, \vu|\eta_0\rangle|^2$ which extends over $a_X$ for $\vr$ and over $a_T$ for $\vu$ in the case of bound state, must be such that $1\simeq a_X^D a_T^D |\langle \vr=0, \vu=0|\eta_0\rangle|^2$. Since $\langle \vp|\vu \rangle=e^{i\vp\cdot \vu}/L^{D/2}$, Eq.~(\ref{WF:rp_ru}) then gives, using again dimensional arguments, $\langle \vr=0, \vp=0|\eta_0\rangle\simeq a_T^{D}/L^{D/2}\sqrt{a_X^{D} a_T^{D}}$; so, we end for the trion ground state with
\be
|\langle \vr=0, \vp=0|\eta_0\rangle|^2\simeq \left(\frac{a_T}{a_XL}\right)^D.\label{DimA:bound1}
\ee
In the case of unbound trions, the $\vu$ extension is of the order of $L$ instead of $a_T$; so, the same dimensional arguments give
\be
|\langle \vr=0, \vp|\eta\rangle|^2\simeq \left(\frac{1}{a_X}\right)^D.\label{DimA:unbound1}
\ee
The $|\langle \vr=0, \vp|\eta\rangle|^2$ function is peaked on various momenta $\vp_\eta$ depending on the unbound trion energies, as understood by noting that, in the absence of electron-exciton interactions, the trion wave function $\langle \vr, \vu|\eta\rangle$ must reduce to a product of an exciton wave function and an electron plane wave,
$\lan\vr,\vu|\eta,S\ran\sim\phi_\nu(\vr)(e^{i\vp_\eta\cdot\vu}/L^{D/2})$. From
 \be
 \lan \vr,\vu|\eta\ran=\sum_{\nu,\vp} \lan \vr|\nu\ran \lan \vu|\vp\ran\lan \nu,\vp|\eta\ran,
 \ee
 with $\lan\vu |\vp\ran=e^{i\vp\cdot\vu}/L^{D/2}$, it is then easy to see that this requires $\lan \nu,\vp|\eta\ran$ to be a delta function $\delta_{\vp_\eta, \vp}$, this delta function being possibly broadened into a peak by electron-exciton interactions.\

We can compute $\langle \vr, \vp|\eta, S\rangle$ from the $\langle \nu,\vp|\eta, S\rangle$ function that we obtain through the numerical resolution of Eq.~(\ref{eq:schroeq_mf1}), via a ``Fourier transform in the exciton sense'', namely
\be
\langle \nu, \vp|\eta, S\rangle=\int  d\vr \langle \nu|\vr \rangle \langle \vr ,\vp|\eta, S\rangle,
\ee
or equivalently
\be
\langle \vr, \vp|\eta, S\rangle=\sum_\nu\langle \vr|\nu \rangle \langle \nu, \vp|\eta, S\rangle.\label{rknS}
\ee
Note that for $\vr=0$, the $\nu$ exciton levels that survive in the $\nu$ sum are $s$-like states only. In practice, in order to compute $\langle \vr, \vp|\eta, S=0\rangle$ from $\langle \nu ,\vp|\eta, S=0\rangle=\phi^{(\mathcal{T},S=0)}_{\vp,i=(\nu,-\vp)}$ with $\mathcal{T}=(\vK=0,\eta)$, as obtained from the resolution of the Schr\"{o}dinger equation (\ref{eq:schroeq_mf1}) along with the parity condition given in Eq.~(\ref{eq:wf_eX2}), we have only kept the three lowest $s$ states in the $\nu$ sum of Eq.~(\ref{rknS}), which is consistent with the fact that we have reduced the electron-exciton basis in this Schr\"{o}dinger equation to these three lowest levels.\

Figure (\ref{fig:WF_rpeta0}) shows $|\langle \vr, \vp|\eta, S=0\rangle|^2$ for the trion bound ground state, $\eta=\eta_0$, and for an unbound state. Note that, due to Eq.~(\ref{DimA:bound1}), we are forced to plot the ground state through $L^2|\langle \vr, \vp|\eta_0, S=0\rangle|^2$ in order to have a quantity independent of sample size $L$. The function $|\langle \vr, \vp|\eta,S=0\rangle|^2$, from now on written as $|\langle \vr, \vp|\eta\rangle|^2$ for simplicity, is concentrated around the origin, $\vr=0,\vp=0$, for bound state trion, whereas it is peaked around a $\vp_\eta$ value for unbound states. (In practice, the $1/L$ width in this figure comes from the momentum discretization used in the numerical resolution). Since the relevant quantity for the trion oscillator strength is going to be $|\langle \vr=0, \vp|\eta\rangle|^2$, we have also given a precise plot of these normalized functions for three unbound states (Fig.~\ref{fig:WF_r=0peta3}). We find that when the unbound trion energy increases, this function broadens due to electron-exciton interactions.\

 \subsection{Trion absorption spectrum}
Knowing the trion wave function for bound and unbound states, it becomes possible to compute the trion absorption spectrum as a function of the photon energy. Trions are usually observed in the optical spectra of lightly doped semiconductor quantum wells. When an exciton is created in a semiconductor through photoabsorption in the presence of an excess of free electrons or free holes, the photocreated exciton interacts with a free carrier to form a trion which can either be in a bound or in an unbound state, depending on the photon energy.   \

Let us first consider an initial state with one free electron with momentum $\vk_i$ and one photon with energy $\omega_{ph}$ and momentum $\vQ_{ph}$. After photon absorption, the final state contains two electrons and one hole, their center-of-mass momentum being $\vk_i+\vQ_{ph}$. Since we are mainly interested in the low-lying trion states, we here focus on the spin singlet $(S=0)$ states. The Fermi golden rule gives the photon absorption as $(-2)$ times the imaginary part of the response function $S^{(T)}(\omega_{ph},\vQ_{ph},\vk_i)$ which, in the case of one photon $(\omega_{ph},\vQ_{ph})$, has been shown to read (see Eq.~(16) of Ref.~\onlinecite{moniqSSC2003_2})
\bea
\lefteqn{S^{(T)}(\omega_{ph},\vk_i)=|\Omega|^2L^D \sum_\eta}\label{ST}\\
&&\times\frac{|\lan \vr=0,\vp_i|\eta\ran|^2}{\omega_{ph}+\frac{\vk_i^2}{2m_e}-\left[\mathcal{E}^{(\eta,S=0)}+\frac{(\vk_i+\vQ_{ph})^2}{2(2m_e+m_h)}\right]+i0^+},\nn
\eea
where $\Omega$ is the vacuum Rabi coupling, while $\vp_i$ is the relative motion momentum of the $(e,X)$ pair. Since the momentum of the photocreated exciton is equal to the photon momentum $\vQ_{ph}$, the total momentum of the $(e,X)$ pair is $\vK_i=\vk_i+\vQ_{ph}$ and its relative motion momentum $\vp_i$ is such that $\vk_i=\vp_i+\beta_e\vK_i$ and $\vQ_{ph}=-\vp_i+\beta_X\vK_i$ with $\beta_e=1-\beta_X=m_e/(2m_e+m_h)$; so $\vp_i=\beta_X\vk_i-\beta_e\vQ_{ph}\simeq\beta_X\vk_i $ as the photon momentum is very small on the characteristic electron scale. By noting that
\be
\frac{\vk_i^2}{2m_e}-\frac{\vk_i^2}{2(2m_e+m_h)}=\frac{(\beta_X\vk_i)^2}{2\mu_{eX}}
\ee
with $\mu_{eX}$ being the effective mass of the electron-exciton pair defined above, we can rewrite Eq.~(\ref{ST}) as
\be
S^{(T)}(\omega_{ph},\vk_i)\simeq|\Omega|^2L^D \sum_\eta\frac{|\lan \vr=0,\beta_X\vk_i|\eta\ran|^2}{\omega_{ph}+\frac{(\beta_X\vk_i)^2}{2\mu_{eX}}-\mathcal{E}^{(\eta,S=0)}+i0^+}.\label{ST2}
\ee
\

Before going further, we wish to make three comments:\

(i) In the absence of electrons in excess in the sample, the absorbed photon creates an exciton, the associated response function reading as
\be
S^{(X)}(\omega_{ph})=|\Omega|^2L^D\sum_\nu \frac{|\lan \vr=0|\nu\ran|^2}{\omega_{ph}-\left[\va^{(\nu)}+\frac{\vQ_{ph}^2}{2(m_e+m_h)}\right]+i0^+}.\label{SX}
\ee
For bound exciton having a spatial extension $a_X$, dimensional arguments lead to a normalized relative motion wave function $\lan \vr|\nu\ran$ such that $1\simeq a_X^D|\lan \vr=0|\nu\ran|^2$; so
\be
|\lan \vr=0|\nu\ran|^2\simeq \left(\frac{1}{a_X}\right)^D.
\ee
It is worth noting that this quantity is of the order of the relative motion wave function for unbound trion states given in Eq.~(\ref{DimA:unbound1}). This is quite reasonable because unbound trions look very much like an exciton with a free electron moving possibly far away from it.\

We then note that Eq.~(\ref{DimA:bound1}) gives an oscillator strength $(a_T/L)^D$ smaller for bound trion than for unbound trion; for large sample size $L$, this {\it a priori} prevents drawing the full absorption spectrum for bound and unbound trions using the same scale. It however is necessary to note that the $L$ factor which here appears, comes from $\lan\vu |\vp\ran=e^{i\vp\cdot\vu}/L^{D/2}$; so, this $L$ factor physically is the length over which the particle keeps its momentum, this length being usually called ``coherence length''. As a result, when the sample volume increases, $(a_T/L)^D$ saturates to $(a_T/L_{coh})^D$ which can be not so small in poor samples. Nevertheless, we will here show the absorption spectra of bound and unbound trion states on different curves to avoid this scaling problem.\

(ii) The $(i0^+)$ term in the response function of an exciton or a trion actually has to be broadened into $i\gamma$ with $\gamma$ being associated with the inverse lifetime of the exciton or trion. Indeed, the response functions quoted in Eqs. (\ref{ST},\ref{SX}) follow from using the Fermi golden rule. This rule is known to be valid for transitions toward a continuum of states. When one photon $\vQ_{ph}$ is absorbed in an undoped semiconductor, we form a well-defined exciton, its momentum being $\vQ_{ph}$. So, the Fermi golden rule should not be used. We have shown (see Ref. \onlinecite{Fubin2005}.) that the Fermi golden rule can still be used for photoabsorption with exciton formation provided that the exciton level is broad enough to be seen as a continuum, the characteristic scale for this broadening $\gamma$ being the vacuum Rabi coupling $\Omega$. When the exciton level is very narrow, as in good quality samples, we say that the system suffers a ``strong coupling'' with the photon field: instead of excitons created by photon absorption through the Fermi golden rule, we have polaritons---the polaritons being the exact eigenstates of one photon coupled to one exciton. So, when using formulas like Eq.~(\ref{ST}) and Eq.~(\ref{SX}) for photon absorption, we implicitly assume the trion or exciton to have a finite lifetime. As a result, $i0^+$ should be replaced by $i\gamma$. \

(iii) The linear response to a field having $N_{ph}$ photons is just $N_{ph}$ times the response function to one photon given in Eq.~(\ref{ST}). This makes the photon absorption linearly increasing with photon number, i.e., laser intensity, as physically required for a linear response. \

Of course, in real experiments, we not only have more than one photon, but also more than one electron in the sample. It turns out that the effect of an electron number increase is far more subtle than a photon number increase because photons are bosons so that they all have the same energy, while electrons are fermions; so, their energies spread out when their number increases. For sample with $N_e$ electrons, cold enough to possibly see trions, we must distinguish two regimes:\

 \begin{figure}[t]
\begin{center}
\epsfig{figure=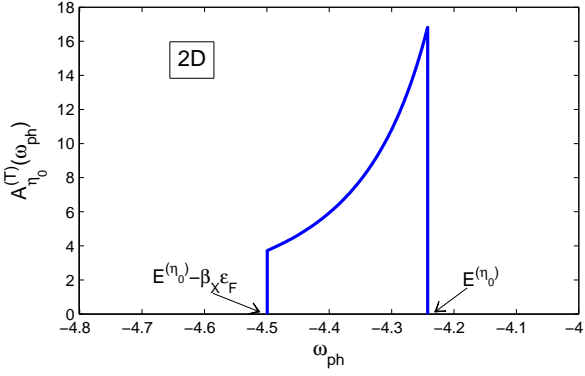,clip=,width=3.4 in}
\end{center}
\caption{(Color online) Absorption $A^{(T)}_{\eta_0}(\omega_{ph})$ (defined in Eq.~(\ref{def:Abs_GS1})) in the presence of cold electrons in 2D quantum wells as a function of the photon energy $\omega_{ph}$ in $R_X^{(3D)}$ unit. The trion ground state energy $\mathcal{E}^{(\eta_0)}$ lies $0.24~R_X^{(3D)}$ below a larger exciton peak which lies at $-4~R_X^{(3D)}$ (not shown). We have taken a mass ratio $m_h/m_e=5$ and set $ N_{ph}|\Omega|^2\rho=1$, the $L^D$ factor in Eq.~(\ref{def:Abs_GS1}) being canceled by the wave function part of this equation, due to Eq.~(\ref{DimA:bound1}). }
\label{fig:result_AbslowT}
\end{figure}

(1) Cold carriers\

For temperature $T$ small compared to the Fermi energy $\va_F$ for these $N_e$ electrons with $\va_F\simeq k_F^2$ and $N_e\simeq (k_FL)^D$, the $N_e$ electrons are degenerate, i.e., they have a probability equal to 1 to have a momentum $\vk_i$ between 0 and $k_F$. So, the response function of these $N_e$ electrons to $N_{ph}$ photons should read, in the absence of interactions, as
\be
S^{(T)}(\omega_{ph})=N_{ph}\sum_{\vk_i=0}^{\vk_F}S^{(T)}(\omega_{ph},\vk_i)\label{ST3}
\ee
with $S^{(T)}(\omega_{ph},\vk_i)$ given in Eq.~(\ref{ST}). By replacing the $\vk_i$ sum by an integral and by setting $\va=(\beta_X\vk_i)^2/2\mu_{eX}=\beta_X\vk_i^2/2m_e$, we get, for 2D systems having a constant density of states $\rho$, the photon absorption spectrum $A_{\eta_0}^{(T)}(\omega_{ph})=-2~{\rm Im}S^{(T)}(\omega_{ph})$ associated with the trion ground state $\eta_0$, through
\bea
\lefteqn{A_{\eta_0}^{(T)}(\omega_{ph})=2\pi N_{ph}|\Omega|^2L^D }\label{def:Abs_GS0}\\
&&\times\int _0^{\beta_X\va_F}\rho d\va\left|\lan \vr=0,\sqrt{2\mu_{eX}\va}|\eta_0\ran\right|^2 \delta(\omega_{ph}+\va-\mathcal{E}^{(\eta_0)})\nn
\eea
which readily gives, for a photon detuning $\delta_{ph}=\mathcal{E}^{(\eta_0)}-\omega_{ph}$ positive,
\be
A_{\eta_0}^{(T)}(\omega_{ph})=2\pi N_{ph}|\Omega|^2L^D\rho \left|\lan \vr=0,\sqrt{2\mu_{eX}\delta_{ph}}|\eta_0\ran\right|^2 \label{def:Abs_GS1}
\ee
for $\mathcal{E}^{(\eta_0)}-\beta_X\va_F<\omega_{ph}<\mathcal{E}^{(\eta_0)}$ and zero otherwise. This shows that $A_{\eta_0}^{(T)}(\omega_{ph})$ essentially follows the decrease of the bound trion relative motion wave function $|\lan \vr=0,\vp|\eta_0\ran|^2$ along $\vp$, the trion absorption spectrum spreading on the low energy side, since more energetic electrons can be used to form the ground state trion (see Fig.~\ref{fig:result_AbslowT}).\

It is worth noting that this absorption spectrum does not increase in amplitude under an electron number increase, but just spreads on the low energy side since $k_F$ increases when $N_e$ increases. Such a spectrum however is rather naive because it forgets many-body effects in the final state similar to the ones leading to Fermi edge singularities: Instead of having a smooth decreasing low-energy tail, these effects bring a peak at the absorption threshold, as discussed and observed in Ref. \onlinecite{moniqEP2005}.\

(2) Hot carriers\

In the case of hot carriers, the $N_e$ dependence of the absorption spectrum is totally different and strongly depends on where the carriers are injected. Let
\be
N(\vk_i,\va_e^*)=N_e\frac{f(\va_{\vk_i},\va_e^*)}{\sum_{\vk_i}f(\va_{\vk_i},\va_e^*)}.\label{eq:N_kT}
\ee
be their density probability. We will here use a Gaussian distribution
\be
f(\va_{\vk_i},\va_e^*)=e^{-\zeta(\vk_i^2/2m_e-\va_e^*)^2},\label{eq:gassian}
\ee
the constants $\zeta$ and $\va_e^*$ being possibly tuned through the bias voltage window of the injected electron current. Summing over the electron momentum $\vk_i$ then amounts to transforming Eq.~(\ref{ST3}) into
\be
S^{(T)}(\omega_{ph},\va_e^*)=N_{ph}\sum_{\vk_i}N(\vk_i,\va_e^*) S^{(T)}(\omega_{ph},\vk_i).\label{ST4}
\ee

\subsubsection{Bound trion}
The absorption spectrum associated with the trion bound state $\eta_0$, given by Eq.~(\ref{def:Abs_GS1}) in the case of cold carriers, transforms into
\bea
A_{\eta_0}^{(T)}(\omega_{ph},\va_e^*)&=&2\pi N_{ph}|\Omega|^2L^D N_e \frac{f(\delta_{ph}/\beta_X,\va_e^*)}{\int^\infty_0 d\va f(\va/\beta_X,\va_e^*)}  \nn\\
&&\times\left|\lan \vr=0,\sqrt{2\mu_{eX}\delta_{ph}}|\eta_0\ran\right|^2. \label{def:Abs_GS2}
\eea
This shows that, for hot carriers, the absorption spectrum is proportional to $N_e$; so, under an electron number increase, the whole spectrum amplitude increases, instead of spreading out as for degenerate electrons. \

\begin{figure}[t]
\begin{center}
\epsfig{figure=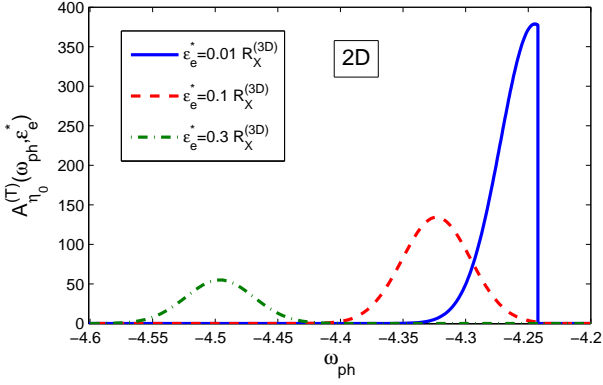,clip=,width=3.4 in}
\end{center}
\caption{(Color online) Absorption $A^{(T)}_{\eta_0}(\omega_{ph},\va_e^*)$ defined in Eq.~(\ref{def:Abs_GS2}) associated with the trion ground state in 2D quantum wells, as a function of the photon energy $\omega_{ph}$ (in $R_X^{(3D)}$ unit) for different parameters $\va_e^*$, and $\zeta=500$ ${R_X^{(3D)}}^{-2}$ in the electron distribution of Eq.~(\ref{eq:gassian}). We have set $N_{ph}|\Omega|^2N_e=1$ and taken a mass ratio $m_h/m_e=5$.}
\label{fig:result_Abs}
\end{figure}

Figure \ref{fig:result_Abs} shows the 2D ground state trion absorption spectrum $A^{(T)}_{\eta_0}(\omega_{ph},\va_e^*)$ for various injected carrier average energies $\va_e^*$ (see Eq.~(\ref{eq:gassian})). This spectrum can be used to probe the hot-carrier distribution. As the photon energy decreases, the absorption intensity follows the decrease of the trion wave function $|\lan \vr=0,\vp_i|\eta\ran|^2$ with relative motion momentum $\vp_i$, as shown in Fig.~\ref{fig:11a}. When the electron number increases, the amplitude of the whole spectrum increases, as a result of the $N_e$ factor in Eq.~(\ref{def:Abs_GS2}), but does not spread, in contrast to the low $T$ limit shown in Fig.~\ref{fig:result_AbslowT}, for which the absorption spreads without changing its amplitude.  \

\subsubsection{Unbound trion}

The situation for unbound trion is somewhat more complex because unbound trion states have energies close to the exciton energies; so, photoabsorption with formation of a trion $A^{(T)}$ is mixed with photoabsorption with formation of an exciton $A^{(X)}$. The resulting absorption spectrum could be thought to read as
\be
fA^{(T)}+(1-f)A^{(X)}=A^{(X)}+f\left[A^{(T)}-A^{(X)}\right],\label{def:Abs_GS3}
\ee
 where $f$ is the capture rate of an electron by an exciton. For $N_e$ electrons in a sample volume $L^D$, this capture rate should be of the order of the exciton volume divided by the average volume occupied by an electron, namely
 \be
 f\simeq \frac{a_X^D}{L^D/N_e}=N_e\left(\frac{a_X}{L}\right)^D.
 \ee
However, electrons with different $\vk_i$'s contribute differently to the trion absorption as seen from Eq.~(\ref{ST}). So, the absorption spectrum should in fact read, instead of Eq.~(\ref{def:Abs_GS3}),
\bea
A(\omega_{ph},\va_e^*)&=&A^{(X)}(\omega_{ph})+\left(\frac{a_X}{L}\right)^D\sum_{\vk_i}N(\vk_i,\va_e^*)\nn\\
&&\times\left[ A^{(T)}(\omega_{ph},\vk_i)-A^{(X)}(\omega_{ph})\right]\label{def:Abs_GS4}
\eea
with $N(\vk_i,\va_e^*)$ given by Eq.~(\ref{eq:N_kT}) for hot carriers. We see that the absorption spectrum reduces to the exciton spectrum in the absence of free carriers, $N_e=0$, as physically required.\

The exciton absorption spectrum $A^{(X)}(\omega_{ph})$ is made of delta peaks centered on the exciton energies $\va^{(\nu)}$, and weighted by the value at $\vr=0$ of the exciton wave function squared $|\lan \vr=0|\nu\ran|^2$ (see Eq.~(\ref{SX})). These delta peaks are broadened when taking into account the finite exciton lifetime, which amounts to replacing $i0^+$ by $i\gamma$ in Eq.~(\ref{SX}) as previously explained; so, the delta functions are then replaced by Lorentzian function with a small half-width $\gamma$, namely
\be
\delta(\omega)\rightarrow \frac{\gamma/\pi}{\omega^2+\gamma^2}.\label{deltaf}
\ee\

Although, according to Eq.~(\ref{def:Abs_GS4}), the photon absorption contains exciton and trion parts, we have chosen here to only show the trion part in order to avoid ambiguity coming from their lifetime yet determined experimentally. The contribution to the photon absorption spectrum resulting from unbound trions is given by
\bea
\lefteqn{A^{(T)}(\omega_{ph},\va_e^*)=N_{ph}\left(\frac{a_X}{L}\right)^D\sum_{\vk_i}N(\vk_i,\va_e^*)A^{(T)}(\omega_{ph},\vk_i)}\nn\\
&=&\frac{2\pi N_{ph}N_e|\Omega|^2a_X^2}{\int^\infty_0 d\va f(\va/\beta_X,\va_e^*)} \int d\va f(\va/\beta_X,\va_e^*)\label{def:Abs_GS_unbound}\\
&&\times \sum_\eta \left|\lan \vr=0,\sqrt{2\mu_{eX}\va}|\eta\ran\right|^2\delta(\omega_{ph}+\va-\mathcal{E}^{(\eta)}),\nn
\eea
the discrete sum over $\eta$ being ultimately replaced by a continuous sum over $\vk_\eta$. To compute it, we have introduced a finite lifetime and replaced the delta function by a Lorentzian function along Eq.~(\ref{deltaf}). This essentially adds a small high-energy tail to the absorption spectrum which would have a sharp edge otherwise.\

\begin{figure}[t]
 \centering \subfigure[] {\ \label{fig:15a}
\includegraphics[width=3.4 in]{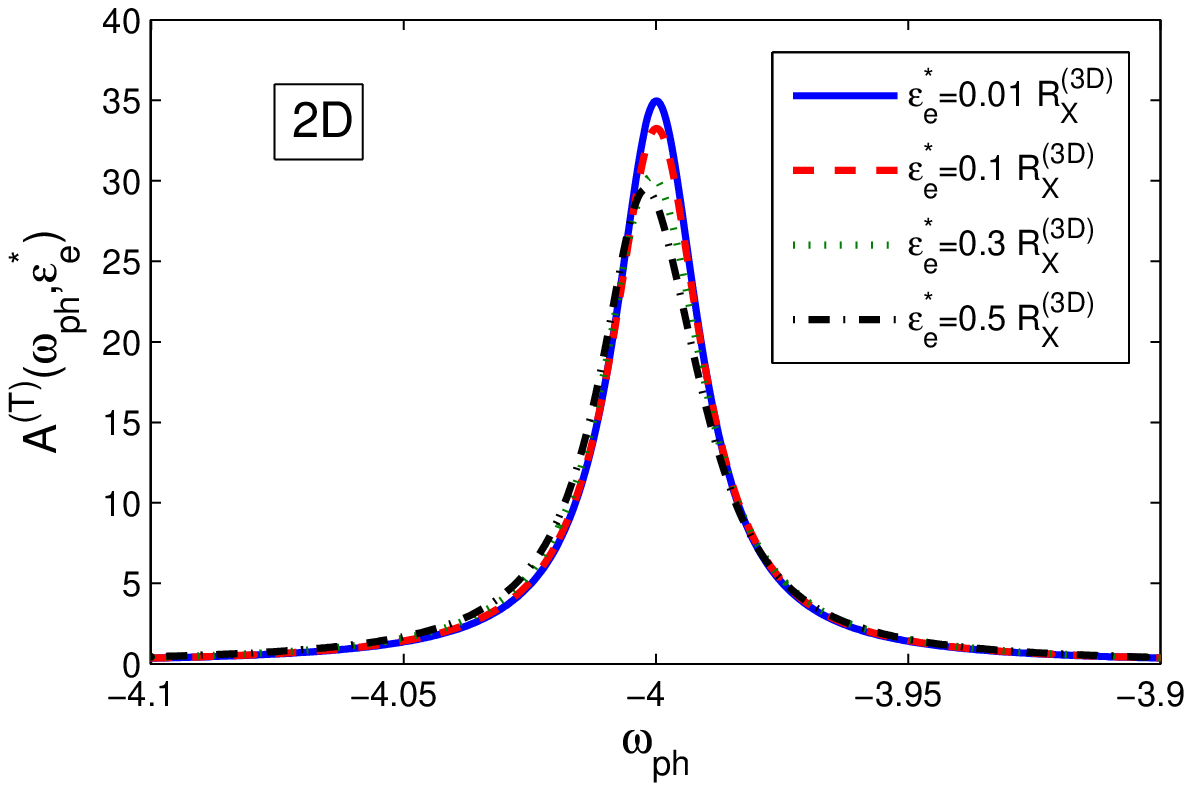} }\\
 \centering \subfigure[] {\ \label{fig:15b}
\includegraphics[width=3.4 in]{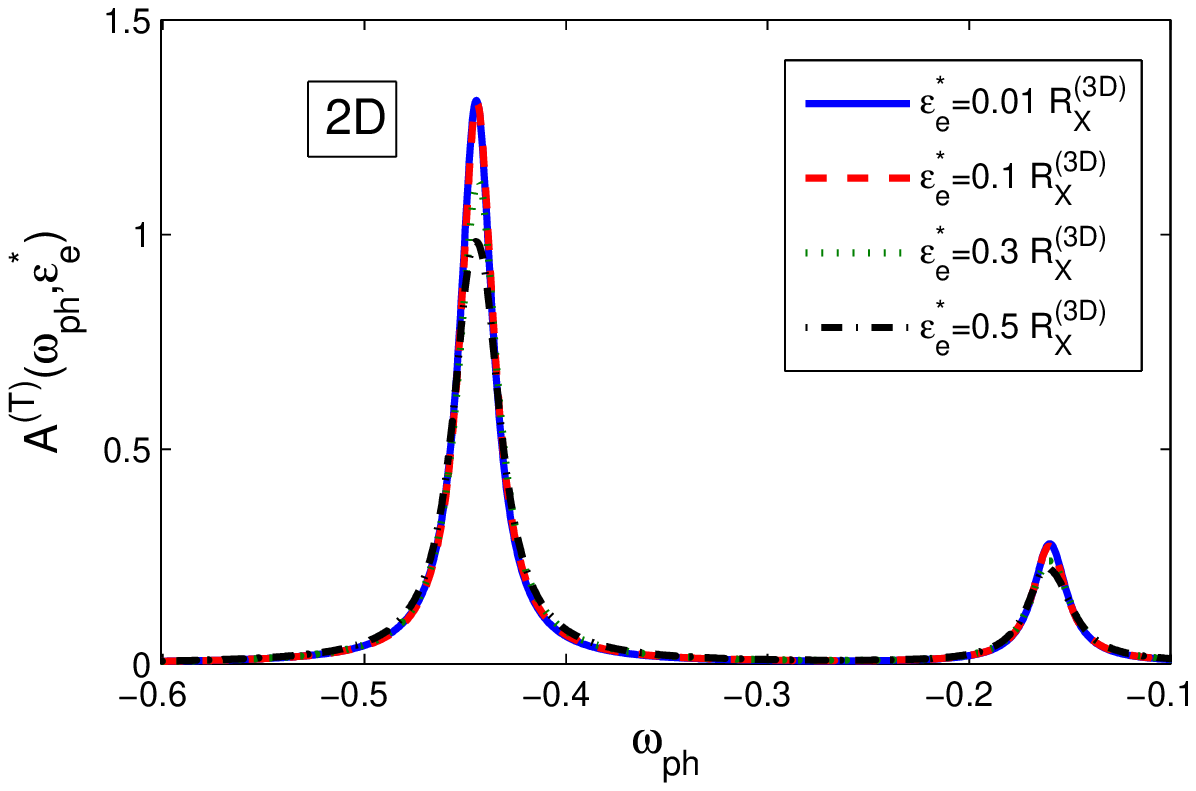}}\\
\caption{(Color online) Unbound trion absorption $A^{(T)}(\omega_{ph},\va_e^*)$ for 2D quantum well as defined in Eq.~(\ref{def:Abs_GS_unbound}) where we have set $ N_{ph}|\Omega|^2N_e=1$. It shows unbound trion peaks resulting from electrons scattering with $1s$ exciton (a) and with $2s$ or $3s$ exciton (b) as a function of the photon energy $\omega_{ph}$ in $R_X^{(3D)}$ unit. We have taken a broadening $\gamma=0.01~R_X^{(3D)}$ in Eq.~(\ref{deltaf}). Other parameters are the same as in Fig.~\ref{fig:result_Abs}.}
\label{fig:result_Abs2D_unbound}
\end{figure}

Figures \ref{fig:15a} and \ref{fig:15b} show the lineshape of the three largest peaks located at the lowest $s$ exciton levels for 2D quantum well, namely $-4~R_X^{(3D)}$, $-4/9~R_X^{(3D)}$, and $-4/25~R_X^{(3D)}$, these peaks being associated with unbound trions, i.e., the electron-exciton scattering states. We see that these peaks spread on both sides of each exciton level, due to energy conservation enforced by the delta function. Their peak height decreases with the increase of the exciton level due to the $\lan \vr=0|\nu\ran$ factor in Eq.~(\ref{rknS}) which for 2D exciton is equal to $4/a_X\sqrt{2\pi(2n-1)^3}$ with $n=(1,2,3)$ for these three peaks.\

The peak lineshape essentially is a Lorentzian which moves slightly toward lower energy side when contributions from electrons with large kinetic energy start to weight in. Moreover, the peak height slightly decreases; this can be attributed to broadening at large relative motion momentum induced by electron-exciton scattering. \

Differences between calculated results and experimental data are expected to primarily come from the finite width of the quantum well. Indeed, all binding energies are known to increase when the space dimension $D$ decreases---as evidenced from the fact that the exciton binding energy is equal to $R_X^{(3D)}$ in 3D but $R_X^{(2D)}=4~R_X^{(3D)}$ in 2D. As a result, the trion binding energy in a finite-width quantum well is expected to be substantially smaller than its exact 2D value. Other sizable effects are expected to come from the confinement of the trion wave function by potential barriers which are not infinite, and also by the potentials resulting from remote donor ions\cite{Eytan1998} or quantum well interface thickness fluctuations\cite{Bracker2005}.  \

\section{Conclusion}
Motivated by the idea that Coulomb interaction between an electron and a hole is far stronger than between an electron and an exciton, we here construct a Schr\"{o}dinger equation for trion using an electron-exciton basis, instead of the standard free carrier basis. One obvious difficulty with the electron-exciton formulation is to exactly handle electron exchange within the electron-exciton pair. This is done through exchange scatterings between electron and exciton similar to the ones which exist between two excitons, as appearing in the composite boson many-body theory. Restricting the exciton basis to the first three $s$-states, we have numerically solved the resulting Schr\"{o}dinger equation for the trion ground state in 2D and 3D systems, as well as for excited bound and unbound states. Using just these three states, our results for the ground state are in reasonable agreement with the ones obtained from the best variational methods. The main advantage of the electron-exciton approach presented here mostly is to reach excited states as easily as the ground state, these excited states being out of reach from usual variational methods. Since we do here solve a Schr\"{o}dinger equation, we are also capable of reaching the trion wave functions of these bound and unbound states. Through these trion wave functions, we can calculate the optical absorption spectrum associated with the trion bound states in the presence of injected hot carriers. The spectra have small peaks associated with the trion bound state and large peaks coming from unbound states, i.e., electron-exciton scattering states.

\section{Acknowledgment}
This work is supported by Academia Sinica, Taiwan. M.C. wishes to thank the Research Center for Applied Sciences of the Academia Sinica in Taiwan for various invitations, while S.-Y. S. wishes to thank C-Nano Ile de France for hospitality at the Institute des NanoSciences de Paris in the summer 2011.

\renewcommand{\thesection}{\mbox{Appendix A}}
\setcounter{section}{0}
\renewcommand{\theequation}{\mbox{A.\arabic{equation}}}
\setcounter{equation}{0} 
\section{Schr\"{o}dinger equation for two electrons and one hole\label{AppendixA}}
We here derive the Schr\"{o}dinger equation (\ref{eq:schrod_diffspin}) starting from Eqs.~(\ref{eq:wf_samespin},\ref{eq:wf_diffspin}). The function $\phi^{(\eta, S)}_{\vk_e \vk_e^\prime \vk_h}$ appearing in this Schr\"{o}dinger equation is linked to the $\psi^{(\eta, S)}_{\vk_e \vk_e^\prime \vk_h}$ prefactors appearing in Eqs.~(\ref{eq:wf_samespin},\ref{eq:wf_diffspin}) through Eq.~(\ref{fac:phi_varphi1}). The interest of this derivation is to evidence that the relevant function is not by no means $\psi^{(\eta, S)}_{\vk_e \vk_e^\prime \vk_h}$ but $\phi^{(\eta, S)}_{\vk_e \vk_e^\prime \vk_h}$.\

We first consider the $(S=1,S_z=2s)$ trion state given in Eq.~(\ref{eq:wf_samespin}). When used into the Schr\"{o}dinger equation (\ref{eq:ham_eeh02}), we get Eq.~(\ref{eq:Schrod_sym}) with $\phi^{(\mathcal{T},S)}_{\vk_e\vk^\prime_e\vk_h}$ replaced by $\psi^{(\mathcal{T},S)}_{\vk_e\vk^\prime_e\vk_h}$. If we now project this equation onto $\lan v| b_{\vp_h}a_{\vp^\prime_e ;s^\prime}a_{\vp_e; s}$, we find two terms that are no longer equal. However, since $V_\vq=V_{-\vq}$, they can be combined to make appear $\left(\psi^{(\mathcal{T},S=1)}_{\vk_e\vk^\prime_e\vk_h}-\psi^{(\mathcal{T},S=1)}_{\vk^\prime_e\vk_e\vk_h}\right)$, which is nothing but $\phi^{(\mathcal{T},S)}_{\vk_e\vk^\prime_e\vk_h}$ for $S=1$. The resulting equation then reduces to Eq.~(\ref{eq:schrod_diffspin}).\

If we now consider the $(S,S_z=0)$ trion states given in Eq.~(\ref{eq:wf_diffspin}), we get the same equation (\ref{eq:Schrod_sym}) with $a^\dag_{\vk_e ;s}a^\dag_{\vk^\prime_e; s^\prime}$ now replaced by $\left[a^\dag_{\vk_e; 1/2}a^\dag_{\vk^\prime_e; -1/2}-(-1)^Sa^\dag_{\vk_e; -1/2}a^\dag_{\vk^\prime_e; 1/2}\right]$ and $\phi^{(\mathcal{T},S)}_{\vk_e\vk^\prime_e\vk_h}$ replaced by $\psi^{(\mathcal{T},S)}_{\vk_e\vk^\prime_e\vk_h}$. When projected onto $\lan v| b_{\vp_h}a_{\vp^\prime_e ;-1/2}a_{\vp_e; 1/2}$, these two terms yield two terms that can again be combined to make appear $\left(\psi^{(\mathcal{T},S)}_{\vk_e\vk^\prime_e\vk_h}+(-1)^S\psi^{(\mathcal{T},S)}_{\vk^\prime_e\vk_e\vk_h}\right)$, so that we also end up with a Schr\"{o}dinger equation for $\phi^{(\mathcal{T},S)}_{\vk_e\vk^\prime_e\vk_h}$ identical to the one given in Eq.~(\ref{eq:schrod_diffspin}).\

\renewcommand{\thesection}{\mbox{Appendix B}} 
\setcounter{section}{0}
\renewcommand{\theequation}{\mbox{B.\arabic{equation}}} 
\setcounter{equation}{0} 

\section{Useful relations on electron-exciton scatterings\label{app:eqs}}
For completeness, this appendix contains the derivation of a few useful relations involving excitons interacting with free electrons, some of which being possibly found in previous works on exciton\cite{moniqPhysReport} and trion\cite{moniqEPJ2003}.

An exciton $i=(\nu_i,\vQ_i)$ with center-of-mass momentum $\vQ_i$, relative motion index $\nu_i$, electron spin $s$ and hole spin $m$, expands on free electron-hole pairs as
\bea
B_{i;s,m}^\dag&=&\sum_{\vk_e\vk_h}a^\dag_{\vk_e;s}b^\dag_{\vk_h;m}\lan \vk_e\vk_h|i\ran\nn\\
&=&\sum_\vp a^\dag_{\vp+\alpha_e\vQ_i ;s}b^\dag_{-\vp+\alpha_h\vQ_i;m}\lan \vp|\nu_i\ran,\label{app:excitonB}
\eea
where $\alpha_e=1-\alpha_h=m_e/(m_e+m_h)$. Conversely, a free electron-hole pair reads in terms of excitons as
\bea
a^\dag_{\vk_e;s}b^\dag_{\vk_h;m}&=&\sum_{i}B_{i;s,m}^\dag\lan i|\vk_e\vk_h\ran\label{app:ehpairs}\\
&=&\sum_{\nu_i}B_{\nu_i,\vk_e+\vk_h; s,m}^\dag\lan \nu_i|\alpha_h\vk_e-\alpha_e\vk_h\ran.\nn
\eea
Since the hole spin $m$ plays no role in problems dealing with one exciton interacting with an electron, we will, from now on, drop the index $m$ from the notations.\

As easy to check from the above equations, electron exchange between one free electron $\vk_e$ with spin $\sigma$ and one exciton $i$ with electron spin $s$ leads to
\bea
B_{i;s}^\dag a^\dag_{\vk_e;\sigma}&=&-\sum_{\vk^\prime_e i^\prime}\lambda\big(\begin{smallmatrix}
\vk_e^\prime& \vk_e\\i^\prime& i\end{smallmatrix}\big)B_{i^\prime; \sigma}^\dag a^\dag_{\vk^\prime_e; s},\label{app:exchange_eX}
\eea
where $\lambda\big(\begin{smallmatrix}
\vk_e^\prime& \vk_e\\i^\prime& i\end{smallmatrix}\big)$ is the Pauli scattering for electron exchange shown in Fig.~\ref{fig:lambda}. It readily follows from its diagrammatic representation that this scattering is given by
\bea
\lambda \big(\begin{smallmatrix}
\vk_e^\prime& \vk_e\\i^\prime& i\end{smallmatrix}\big)&=&\sum_{\vp_h} \lan i^\prime|\vk_e\vp_h\ran\lan \vp_h\vk^\prime_e|i\ran.\label{app:func_lambda}
\eea
This compact expression of the Pauli scattering between $(e,X)$ pairs allows us to check that two exchanges reduce to an identity, as a result of the closure relation on $|i\ran$ exciton states
\bea
\lefteqn{\sum_{i^{\prime\prime},\vp_e }\lambda\big(\begin{smallmatrix}
\vk_e^\prime& \vp_e\\i^\prime& i^{\prime\prime}\end{smallmatrix}\big)\lambda\big(\begin{smallmatrix}
\vp_e&\vk_e \\i^{\prime\prime}& i\end{smallmatrix}\big)}\nn\\
&&=\sum_{ i^{\prime\prime},\vp_e,\vq_h,\vq_h^\prime }\lan i^\prime|\vp_e\vq^\prime_h\ran \lan \vq^\prime_h\vk^\prime_e|i^{\prime\prime}\ran \lan i^{\prime\prime}|\vk_e\vq_h\ran \lan \vq_h\vp_e|i\ran\nn\\
&&=\delta_{i^\prime i}\delta_{\vk^\prime_e\vk_e}.\label{app:lambdaIden}
\eea
As seen from Fig.~\ref{fig:lambda2}, for an exciton labeled by $i=(\nu_i,\vQ_i)$ this Pauli scattering reduces to $\sum \lan \nu_{i^\prime}|\vp^\prime\ran \lan \vp|\nu_{i}\ran$ provided that the $(\vp,\vp^\prime)$ momenta are such that $\vk^\prime_e=\vp+\alpha_e\vQ_i,$ $\vk_e=\vp^\prime+\alpha_e\vQ_{i^\prime}$ and $-\vp^\prime+\alpha_h\vQ_{i^\prime}=-\vp+\alpha_h\vQ_i$. This gives
\be
\lambda\big(\begin{smallmatrix}
\vk_e^\prime& \vk_e\\
i^\prime& i
\end{smallmatrix}\big)=\delta_{\vk^\prime_e+\vQ_{i^\prime},\vk_e+\vQ_i}\lan \nu_{i^\prime}|\vk_e-\alpha_e\vQ_{i^\prime}\ran\lan \vk^\prime_e-\alpha_e\vQ_i|\nu_i\ran.\label{app:func_lambda2}
\ee

These Pauli scatterings formally appear through two commutators
\bea
\left[B_{{i^\prime};s^\prime},B^\dag_{i;s}\right]&=&\delta_{{i^\prime} i}\delta_{s^\prime s}-D_{i^\prime s^\prime;is},\label{app:commu_BB}\\
\left[D_{{i^\prime} s^\prime;is},a_{\vk_e;\sigma}^\dag\right]&=& \delta_{s^\prime \sigma}\sum_{\vk_e^\prime} \lambda\big(\begin{smallmatrix}
\vk_e^\prime& \vk_e\\ {i^\prime}& i\end{smallmatrix}\big)a_{\vk^\prime_e;s}^\dag.\label{app:commu_D}
\eea
From them, it is easy to show that the scalar product of two electron-exciton pairs is given by
\bea
&&\lan v|a_{\vp_e;\sigma}B_{j;\sigma^\prime} B^\dag_{i;s^\prime}a_{\vk_e; s}^\dag|v\ran\nn\\
&&=\lan v|a_{\vp_e;\sigma}\Big\{\left[B_{j;\sigma^\prime} ,B^\dag_{i;s^\prime}\right]+B^\dag_{i;s^\prime}B_{j;\sigma^\prime}\Big\}a_{\vk_e; s}^\dag|v\ran\nn\\
&&= \delta_{\sigma s}\delta_{\sigma^\prime s^\prime}\delta_{\vp_e\vk_e}\delta_{ji}-\delta_{\sigma s^\prime}\delta_{\sigma^\prime s}\lambda \big(\begin{smallmatrix}
\vp_e& \vk_e\\j& i\end{smallmatrix}\big).\label{app:scalarprod_eX}
\eea

Exciton and electron also interact by direct Coulomb scattering $\xi\big(\begin{smallmatrix}
\vk_e^\prime& \vk_e\\{i^\prime}& i\end{smallmatrix}\big)$ which also follows from two commutators. For an electron-hole Hamiltonian $H$, they appear through
\bea
\left[H,B^\dag_{i;s}\right]&=&E_i^{(X)}B^\dag_{i;s}+V_{i;s}^\dag,\label{app:commu_B}\\
\left[V_{i;s}^\dag,a_{\vk_e; s^\prime}^\dag\right]&=&\sum_{\vk^\prime_e {i^\prime}} \xi\big(\begin{smallmatrix}
\vk_e^\prime& \vk_e\\{i^\prime}& i\end{smallmatrix}\big)B^\dag_{{i^\prime};s}a_{\vk_e^\prime ;s^\prime}^\dag.\label{app:commu_V}
\eea
Its diagrammatic representation, shown in Fig.~\ref{fig:Xi}, readily gives
\bea
\xi\big(\begin{smallmatrix}
\vk_e^\prime& \vk_e\\{i^\prime}& i\end{smallmatrix}\big)&=&V_{\vk^\prime_e-\vk_e}\sum_{\vp_e\vp_h}\Big[\lan {i^\prime}|\vp_e+\vk_e-\vk_e^\prime,\vp_h\ran\nn\\
&&-\lan {i^\prime}| \vp_e,\vp_h+\vk_e-\vk_e^\prime\ran\Big]\lan \vp_h\vp_e|i\ran.\label{app:direcCoulomb00}
\eea
For $i=(\nu_i,\vQ_i)$, this scattering reduces to $\sum \lan \nu_{i^\prime}|\vp^\prime\ran \lan \vp|\nu_{i}\ran$ provided that the $(\vp,\vp^\prime)$ momenta are such that $\vp^\prime+\alpha_e\vQ_{i^\prime}=\vp+\alpha_e\vQ_i+\vk_e-\vk_e^\prime$ and $-\vp^\prime+\alpha_h\vQ_{i^\prime}=-\vp+\alpha_h\vQ_i$. This gives
\bea
\xi\big(\begin{smallmatrix}
\vk_e^\prime& \vk_e\\{i^\prime}& i\end{smallmatrix}\big)&=&\delta_{\vQ_{i^\prime}+\vk_e^\prime,\vQ_i+\vk_e}V_{\vk^\prime_e-\vk_e}\sum_{\vp}\Big[\lan \nu_{i^\prime}|\vp+\alpha_h(\vk_e-\vk_e^\prime)\ran\nn\\
&&-(\alpha_h\rightarrow-\alpha_e)\Big]\lan\vp|\nu_i\ran.\label{app:direcCoulomb01}
\eea

We can mix the direct Coulomb scattering $\xi$ with the electron exchange $\lambda$ to get the ``in'' and ``out'' exchange-Coulomb scatterings defined as
\bea
\xi^{\rm in}\big(\begin{smallmatrix}
\vk_e^\prime& \vk_e\\i^\prime& i\end{smallmatrix}\big)&=&\sum_{\vk^{\prime\prime}_e i^{\prime\prime}}\lambda\big(\begin{smallmatrix}
\vk_e^\prime& \vk^{\prime\prime}_e\\i^\prime& i^{\prime\prime}\end{smallmatrix}\big)\xi\big(\begin{smallmatrix}
\vk_e^{\prime\prime}& \vk_e\\i^{\prime\prime}& i\end{smallmatrix}\big)\label{eq:xi_in},\\
\xi^{\rm out}\big(\begin{smallmatrix}
\vk_e^\prime& \vk_e\\i^\prime& i\end{smallmatrix}\big)&=&\sum_{\vk^{\prime\prime}_e i^{\prime\prime}}\xi\big(\begin{smallmatrix}
\vk_e^\prime& \vk^{\prime\prime}_e\\i^\prime& i^{\prime\prime}\end{smallmatrix}\big)\lambda\big(\begin{smallmatrix}
\vk_e^{\prime\prime}& \vk_e\\i^{\prime\prime}& i\end{smallmatrix}\big).\label{eq:xi_out}
\eea
Using Fig.~\ref{fig:Xi_in}, we find that the ``in'' exchange scattering is given by
\bea
\xi^{\rm in}\big(\begin{smallmatrix}
\vk_e^\prime& \vk_e\\{i^\prime}& i\end{smallmatrix}\big)&=&\sum_{\vq\not=0}V_{\vq}\sum_{\vp_e\vp_h}\Big[\lan {i^\prime}|\vk_e-\vq,\vp_h\ran\label{app:inexCoulomb01}\\
&&-\lan {i^\prime}| \vk_e-\vq,\vp_h+\vq\ran\Big]\lan \vp_h\vp_e|i\ran.\nn
\eea
If we now write the exciton $i$ as $(\nu_i,\vQ_i)$, Figure \ref{fig:Xi_in2} gives this ``in'' exchange scattering as
$\sum V_{\vq} \lan \nu_{i^\prime}|\vp^\prime\ran \lan \vp|\nu_{i}\ran$ provided that the $(\vp,\vp^\prime)$ momenta are such that $\vk_e^\prime=\vp+\alpha_e\vQ_i+\vq,$ $\vp^\prime+\alpha_e\vQ_{i^\prime}=\vk_e-\vq,$ and $-\vp^\prime+\alpha_h\vQ_{i^\prime}=-\vp+\alpha_h\vQ_i$ for the electron-electron part, while we must have $\vk_e^\prime=\vp+\alpha_e\vQ_i,$ $\vp^\prime+\alpha_e\vQ_{i^\prime}=\vk_e-\vq,$ and $-\vp^\prime+\alpha_h\vQ_{i^\prime}=-\vp+\alpha_h\vQ_i+\vq$ for the electron-hole part. This ultimately leads to
\bea
\lefteqn{\xi^{\rm in}\big(\begin{smallmatrix}
\vk_e^\prime& \vk_e\\{i^\prime}& i\end{smallmatrix}\big)=\delta_{\vQ_{i^\prime}+\vk_e^\prime,\vQ_i+\vk_e}\sum_\vq V_{\vq}\lan \nu_{i^\prime}| \vk_e-\vq-\alpha_e \vQ_{i^\prime}\ran}\nn\\
&&\times\Big[\lan \vk^\prime_e-\vq-\alpha_e\vQ_i|\nu_i\ran-\lan \vk_e^\prime-\alpha_e\vQ_i |\nu_i\ran\Big].\label{app:inexCoulomb02}
\eea
These two exchange-Coulomb scatterings are related to the dimensionless Pauli scattering for electron exchange through
\bd
(E_{\vk_e^\prime i^\prime}-E_{\vk_e i})\lambda\big(\begin{smallmatrix}
\vk_e^\prime& \vk_e\\i^\prime& i\end{smallmatrix}\big)=\xi^{\rm in}\big(\begin{smallmatrix}
\vk_e^\prime& \vk_e\\i^\prime& i\end{smallmatrix}\big)-\xi^{\rm out}\big(\begin{smallmatrix}
\vk_e^\prime& \vk_e\\i^\prime& i\end{smallmatrix}\big),\label{eq:rel_Pauliexchange}
\ed
as easy to show by calculating $\lan v|a_{\vk_e^\prime; s}B_{i^\prime; s}H B^\dag_{i; s}a^\dag_{\vk_e; s}|v\ran$ with $H$ acting either on the left or on the right.

\end{document}